\documentclass[12pt]{article}
\usepackage{t1enc}
\usepackage{times}
\usepackage{fullpage}     
\usepackage{amsfonts}

\newcommand{\set}[1]{\left\{ #1\right\}}
\newcommand{\gilt}{:}
\newcommand{\sodass}{\,:\,}
\newcommand{\setGilt}[2]{\left\{ #1\sodass #2\right\}}

\newcommand{\realrange}[2]{\left[#1, #2\right]}

\newcommand{\unitrange}[2]{\realrange{0}{1}}

\newcommand{\llabel}[1]{\label{\labelprefix:#1}}
\newcommand{\labelprefix}{} 

\newcommand{\discussionsize}{\small}

\newcommand{\frage}[1]{}

\newenvironment{code}{\noindent
\begin{tabbing}%
\hspace{2em}\=\hspace{2em}\=\hspace{2em}\=\hspace{2em}\=\hspace{2em}\=%
\hspace{2em}\=\hspace{2em}\=\hspace{2em}\=\hspace{2em}\=\hspace{2em}\=%
\kill}{\end{tabbing}}

\newcommand{\labelcommand}{}
\newcommand{\captiontext}{}
\newsavebox{\codeparam}
\newcounter{lineNumber}
\newenvironment{disscodepos}[3]{%
\renewcommand{\labelcommand}{#2}%
\renewcommand{\captiontext}{#3}%
\sbox{\codeparam}{\parbox{\textwidth}{#3}}%
\begin{figure}[#1]\begin{center}\begin{code}\setcounter{lineNumber}{1}}{%
\end{code}\end{center}\caption{\llabel{\labelcommand}\captiontext}\end{figure}}

{\end{disscodepos}}

\newcommand{\Is}       {:=}

\newdimen\endofsize\endofsize=0.5em
\def\endofbeweis{~\quad\hglue\hsize minus\hsize
                 \hbox{\vrule height \endofsize width
\endofsize}\par}

\usepackage{epsfig}
\usepackage{url}
\usepackage{amsmath}
\usepackage{amssymb}
\usepackage{pdflscape}
\usepackage{latexsym}

\def\MdR{\ensuremath{\mathbb{R}}}

\newcounter{mybibitemcount}

                         {\end{list}}
\newcommand{\innerOuter}{\mathrm{innerOuter}}
\newcommand{\expansion}{\mathrm{expansion}}
\newcommand{\algname}{KaPPa}
\newcommand{\Color}{\mathcal{C}}
\newcommand{\Id}[1]{\ensuremath{\mathit{#1}}}

\newenvironment{proof}{{\bf Proof: }}{}
\newcommand{\mytitle}{Engineering a Scalable High Quality Graph Partitioner}
\title{\mytitle}
\author{Manuel Holtgrewe, Peter Sanders, Christian Schulz}
\date{}
\begin{document}
\maketitle
\begin{abstract}
  We describe an approach to parallel graph partitioning that scales to hundreds
  of processors and produces a high solution quality. 
  For example, for many instances from Walshaw's benchmark collection
  we improve the best known partitioning.
  We use the well known
  framework of multi-level graph partitioning. All components are implemented by
  scalable parallel algorithms. Quality improvements compared to previous
  systems are due to better prioritization of edges to be contracted, better
  approximation algorithms for identifying matchings, better local search
  heuristics, and perhaps most notably, a parallelization of the FM local
  search algorithm that works more locally than previous approaches.
\end{abstract}

\section{Introduction}

Many important applications of computer science involve processing large graphs,
e.g., stemming from finite element methods, digital circuit design, route
planning, social networks,\ldots Very often these graphs need to be partitioned
or clustered such that there are few edges between the blocks (pieces).  In
particular, when you process a graph in parallel on $k$ PEs (processing
elements) you often want to partition the graph into $k$ blocks of about equal
size.  In this paper we focus on a version of the problem that constrains the
maximum block size to $(1+\epsilon)$ times the average block size and tries to
minimize the total cut size, i.e., the number of edges that run between blocks.
It is well known that there are more realistic (and more complicated) objective
functions involving also the block that is worst and the number of its
neighboring nodes \cite{Hendrickson98} but minimizing the cut size has been adopted as
a kind of standard since it is usually highly correlated with the other
formulations. We believe that the results presented here will be adaptable to
other objective functions and also to other setting such as graph
clustering where $k$ and the block sizes are not necessarily fixed.

We begin in Section~\ref{s:preliminaries} by introducing basic concepts.
The main part of the paper are the sections on contraction \ref{s:contraction},
initial partitioning \ref{s:initial}, and refinement \ref{s:refinement}.
Section~\ref{s:experiments} summarizes extensive experiments done
to tune the algorithm and evaluate its performance.
Some related work is discussed in Section~\ref{s:related} and
Section~\ref{s:conclusions} summarizes the results and gives some outlook on 
future work.

\section{Preliminaries}\label{s:preliminaries}

Consider an undirected graph $G=(V,E,c,\omega)$ 
with edge weights $\omega: E \to \MdR_{>0}$, node weights
$c: V \to \MdR_{\geq 0}$, $n = |V|$, and $m = |E|$.
We extend $c$ and $\omega$ to sets, i.e.,
$c(V')\Is \sum_{v\in V'}c(v)$ and $\omega(E')\Is \sum_{e\in E'}\omega(e)$.
$\Gamma(v)\Is \setGilt{u}{\set{v,u}\in E}$ denotes the neighbors of $v$.

We are looking for \emph{blocks} of nodes $V_1$,\ldots,$V_k$ 
that partition $V$, i.e., $V_1\cup\cdots\cup V_k=V$ and $V_i\cap V_j=\emptyset$
for $i\neq j$. The \emph{balancing constraint} demands that 
$\forall i\in 1..k\gilt c(V_i)\leq L_{\max}\Is (1+\epsilon)c(V)/k+\max_{v\in V} c(v)$ for
some parameter $\epsilon$.
The objective is to minimize the total \emph{cut} $\sum_{i<j}w(E_{ij})$ where 
$E_{ij}\Is\setGilt{\set{u,v}\in E}{u\in V_i,v\in V_j}$. 
By default, our initial inputs will have unit edge and node weights. 
However, even those will be translated into weighted problems in the course of the algorithm.

A matching $M\subseteq E$ is a set of edges that do not share any common nodes,
i.e., the graph $(V,M)$ has maximum degree one.  

An edge coloring $\Color$ assigns a color (a number)
to each edge of a graph such that no two incident edges have the same color. Note
that the edges with a particular color define a matching, i.e., $\Color$
partitions the edges into matchings. We will be interested in colorings with
a small number of different colors used. 

\emph{Contracting} an edge
$\set{u,v}$ means to replace the nodes $u$ and $v$ by a new node $x$ connected
to the former neighbors of $u$ and $v$. We set $c(x)=c(u)+c(v)$. If replacing
edges of the form $\set{u,w},\set{v,w}$ would generate two parallel edges
$\set{x,w}$, we insert a single edge with
$\omega(\set{x,w})=\omega(\set{u,w})+\omega(\set{v,w})$.
\emph{Uncontracting} an edge $e$ undos its contraction. 
In order to avoid tedious notation, $G$ will denote the current state of the graph
before and after a (un)contraction unless we explicitly want to refer to 
different states of the graph.

The multilevel approach to clustering consists of three main phases.

In the \emph{contraction} (coarsening) phase, 
we iteratively identify matchings $M\subseteq E$ 
and contract the edges in $M$. This is repeated until $|V|$ falls below some threshold.
Contraction should quickly reduce the size of the input and each computed level
should be reflect the global structure of the input network. In particular,
nodes should represent densely connected subgraphs.

Contraction is stopped when the graph is small enough to be directly
partitioned in the \emph{initial partitioning phase} using some other algorithm. 
We could actually use a trivial initial
partitioning algorithm if we contract until exactly $k$ nodes are left. However,
if $|V|\gg k$ we can afford to run some fairly expensive algorithm for initial
partitioning.

In the \emph{refinement} (or uncoarsening) phase, the matchings are iteratively
uncontracted.  After uncontracting a matching, the refinement algorithm moves
nodes between blocks  in order to reduce the cut size or balance.  The nodes to
move are often found using some kind of local search. The intuition
behind this approach is that a good partition at one level of the hierarchy
will also be a good partition on the next finer level so that refinement will
quickly find a good solution.

\section{Contraction}\label{s:contraction}

We distinguish two separate choices for computing a matching: A \emph{rating
  function} for the edges telling us which edges are how valuable for the
matching and a \emph{matching algorithm} that tries to find a matching of near
maximum weight efficiently. Contractions are run until
the graph is ``small enough''.

\subsection{Edge Rating}

In most previous work,
the edge weight $\omega(e)$ itself is used as a rating function (see Section~\ref{s:related} for more details). We additionally consider

\newcommand{\Outer}{\mathrm{Out}}
$$\expansion(\set{u,v})\Is \frac{\omega(\set{u,v})}{c(u)+c(v)}$$
$$\expansion^*(\set{u,v})\Is \frac{\omega(\set{u,v})}{c(u)c(v)}$$
$$\expansion^{*2}(\set{u,v})\Is \frac{\omega(\set{u,v})^2}{c(u)c(v)}$$
$$\innerOuter(\set{u,v})\Is \frac{\omega(\set{u,v})}{\Outer(v)+\Outer(u)-2\omega({u,v})}$$

where $\Outer(v)\Is \sum_{x\in\Gamma(v)}\omega(\set{v,x})$.
These bounds  are heuristically inferred from a few basic
principles: its good to contract heavy edges because this decreases the cut size. 
For the same reason we want to avoid clusters with many outgoing edges.
Furthermore, we preferably contract light nodes because we want to keep the node weight
at any level of contraction reasonably uniform. 

In \cite{DAHoltgrewe} several other functions based on ratings used in graph clustering are considered.
However, they did not lead to very good results so that we do not go into details here.

\subsection{Sequential Matching Algorithms}\label{ss:seqMatching}
Although the maximum weight matching problem can be solved optimally in polynomial time,
the available algorithms are too slow for very large graphs so that all 
graph partitioners use fast approximation algorithms.
We tried three different matching algorithms that all run in linear or near linear time:

\noindent{\bf SHEM:} \emph{Sorted Heavy Edge Matching} is the algorithm used in Metis \cite{SchKarKum00}.
  The nodes are sorted by increasing degree and then scanned. For each scanned node $v$,
  the heaviest edge $\set{u,v}$ incident to $v$ is put into the matching and all remaining 
  edges incident to $u$ and $v$ are excluded from further consideration.
  This algorithm is very fast but cannot give any worst case guarantees.\\[1mm]
\noindent{\bf Greedy:} The edges are sorted by descending weight and then scanned.
  When edge $\set{u,v}$ and neither $u$ nor $v$ are matched yet, $\set{u,v}$
  is put into the matching. The Greedy algorithm guarantees a matching whose weight
  is at least half of the weight of a maximum weight matching.\\[1mm]
\noindent{\bf GPA:} The \emph{Global Path Algorithm} was proposed in \cite{MauSan07} as
  a synthesis of the Greedy algorithm and the Path Growing Algorithm \cite{DH03a}.
  All three algorithms achieve a half-approximation in the worst case, but empirically,
  GPA gives considerably better results. Similar to Greedy, GPA 
  scans the edges in order of decreasing weight
  but rather than immediately building a matching, it first constructs a collection
  of paths and even cycles. Afterwards, optimal solutions are computed for each
  of these paths and cycles using dynamic programming. 

We have not tried more sophisticated linear time algorithms that
achieve 2/3-approximations since in \cite{MauSan07}
they empirically turn out to be much slower yet not much better than GPA.

\subsection{Parallel Matching Algorithms}

In our basic strategy we follow \cite{ManneB07}.
We first compute a preliminary partition of the graph, e.g., using
coordinate information. Currently we have implemented a recursive bisection
algorithm for nodes with 2D coordinates that alternately splits the data by the
$x$-coordinate and the $y$-coordinate \cite{Bentley75,BergerB85}. We can also use the initial numbering of
the nodes. Note that the initial partitioning does not directly affect the
final partitioning computed later -- its main purpose is to increase locality for
the compuation of matchings. 

We then combine a sequential matching algorithm running on each
partition and a parallel matching algorithm running on the \emph{gap graph}.
The gap graph consists on those edges $\set{u,v}$ where $u$ and $v$ reside on different
PEs and $\omega(\set{u,v})$ exceeds the weight of the edges that may have been matched
by the local matching algorithms to $u$ and $v$. The parallel matching algorithm itself
iteratively matches edges that $\set{u,v}$ are locally heaviest both at $u$ and $v$ until
no more edges can be matched.

\section{Initial Partitioning}\label{s:initial}

\sloppypar The contraction is stopped when the number of remaining nodes on some PE is
below $\max{(20,n/(\alpha k^2))}$ for some tuning parameter $\alpha$. The graph is then small
enough to be partitioned on a single PE. 
Our framework allows using pMetis or Scotch for initial
partitioning. We use the sequential algorithms and run them simultaneously on
all PEs, each with a different seed for the random number generator.
Since initial partitioning is very fast, it is also repeated several times.
The best solution is then broadcast to all PEs. 

\section{Refinement}\label{s:refinement}
\begin{figure}[b]
\begin{center}
\includegraphics[width=150pt]{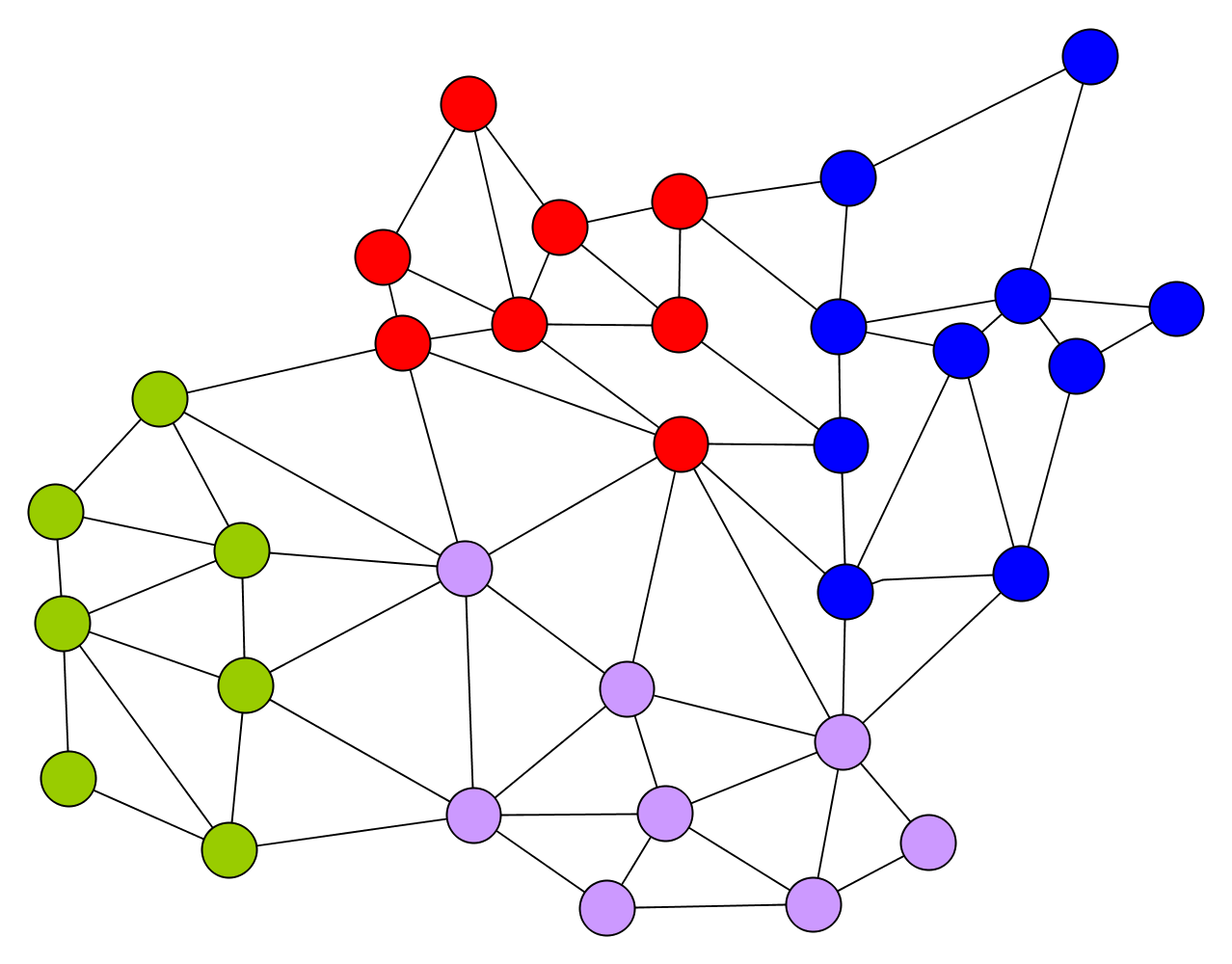}{\Large$\Rightarrow$}\includegraphics[width=150pt]{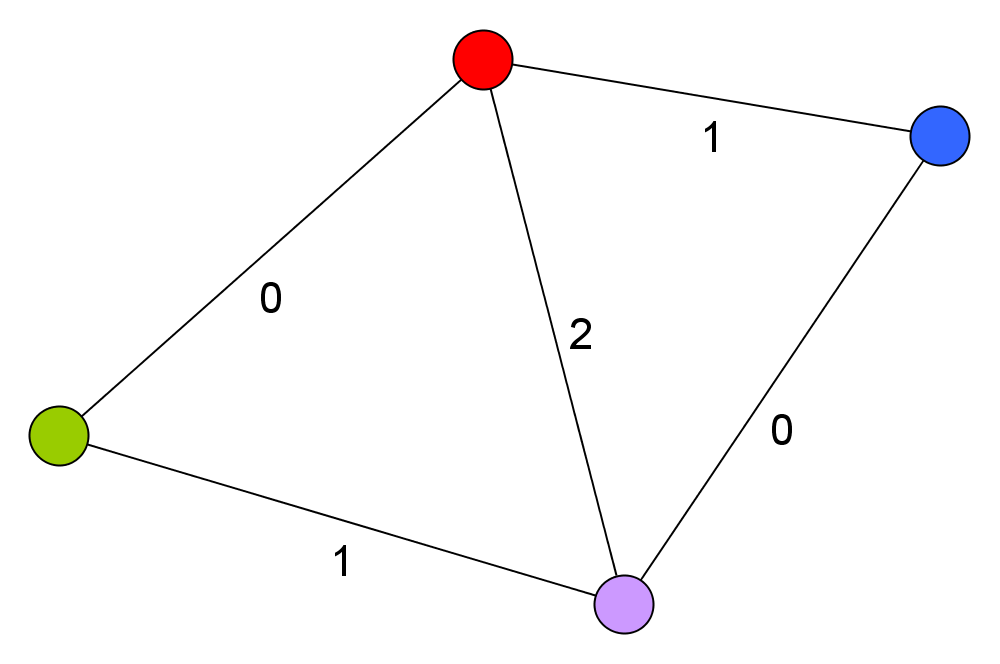}
\end{center}
\vspace*{-8mm}
\caption{A graph which is partitioned into four blocks and its corresponding
  quotient graph $\mathcal{Q}$.  The quotient graph has an edge coloring
  indicated by the numbers and each edge set induced by edges with the same
  color form a matching $\mathcal{M}(c)$. Pairs of blocks with the same color
  can be refined in parallel.}
\label{fig:pictureoverview}
\end{figure}

Recall that the refinement phase iteratively uncontracts the matchings
contracted during the contraction phase.  After a matching is uncontracted,
local search based refinement algorithms move nodes between block boundaries in
order to reduce the cut while maintaining the balancing constraint. As most
other current systems, we adopt the basic approach from \cite{fiduccia1982lth}
which runs in linear time.  The basic idea behind our parallel refinement
algorithm is that at any time, each PE may work on one pair of neighboring
blocks performing a local search constrained to moving nodes between these two
blocks.  In order to assign pairs of blocks to PEs, we use the \emph{quotient
  graph} $Q$ whose nodes are blocks of the current partition and whose edges
indicate that there are edges between these blocks in the underlying graph $G$.
Since we have the same number of PEs and blocks, each PE will work the block
assigned to it and at one of its neighbors in $Q$. From now on, we will
therefore identify blocks and PEs.  Figure~\ref{fig:pictureoverview} gives an example.

We use matchings of $Q$ to define with which neighbor in $Q$ a PE is
working at a particular point in time. If ${u,v}$ is in the matching,
both corresponding PEs will refine the partitions $u$ and $v$ using
different seeds for their random number generator. See Section~\ref{ss:twoblocks}
for more details.
After the local search is finished, the better partitioning of the 
two blocks is adopted. 

Of course, for a good partition, we need to perform local search on every edge
of $Q$ eventually (we call this a \emph{global iteration}). Section~\ref{ss:edgecoloring} describes our approaches for
ensuring this.

We ensure this by iterating through the matchings defined
by an edge coloring of $Q$.  See Section~\ref{ss:edgecoloring}
for more details.

Overall, this approach naturally defines a nested loop controlling our local
search strategy. The innermost loop moves nodes between two blocks using the FM-algorithm \cite{fiduccia1982lth}. A \emph{local iteration} repeats this local search. 
A \emph{global iteration} iterates over the colors of an edge coloring.  
The loops terminate when either no improvement was found 
(in strong variants: when no improvement was found twice in a row.)
or
when a preset maximum number of iterations is exceeded.

\subsection{Choosing Matchings}\label{ss:edgecoloring}

We have implemented two strategies. One finds edges of $Q$ not yet used
for local search in a randomized local way. The other steps through the colors of 
an edge coloring of the quotient graph $Q$. Note that this requires only
local synchronization between PEs actually collaborating at a particular point in time.
We only describe the latter one here since it performs slightly better in
our experiments.
Our coloring algorithm
is a parallelization of a well known
sequential greedy edge coloring algorithm: Each PE has a set $\mathcal{L}$ of free
colors that have not been used for coloring incident edges. In each round
of the algorithm, PEs throw a coin with sides \Id{active} and \Id{passive}. An
active PE $u$ picks a random incident uncolored edge $\set{u,v}$ and sends this
edge together with its free-list to PE $v$.  These \emph{requests} are rejected
if they are sent to other active PEs.  Passive PEs $v$ process requests
$(\set{u,v},\mathcal{L'})$ by choosing the color $c=\min L\cap L'$ for edge
$\set{u,v}$ and sending $c$ back to $u$.
This algorithm is repeated until all edges are colored. It can be shown that this
algorithm needs at most twice as many colors as an optimal edge coloring.

\subsection{Refinement Between Two Blocks}\label{ss:twoblocks}
\begin{figure}
\begin{center}
\includegraphics[width=\textwidth]{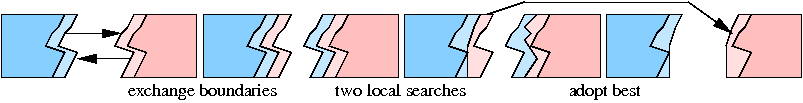}
\end{center}
\vspace{-7mm}
\caption{Refinement between two blocks using boundary exchange.}
\label{fig:refinepair}
\end{figure}

We use a fully distributed graph data structure.  More precisely, we use hybrid
between a static and a dynamic graph data structure.  Immediately after
uncontracting a matching, every PE stores the partition it is responsible for in
a static adjacency array representation (also called forward-star
representation), i.e., there is an edge array storing target nodes and edge
weights and a node array storing node weights and the start of the relevant
segment in the edge array. In addition, we use a hash table to store migrated
nodes and a second edge array for the corresponding edges.  See \cite{DASchulz}
for more details. Before a local search operation, we perform a bounded breadth
first search starting from the boundary of each block, and send copies of
this boundary array to the partner PE in the local search. The local search is
then limited to this boundary area. 
This way, for large
graphs, only a small fraction of each block has to be communicated. If it should
really happen that the local search would profit from going beyond the boundary
area, this will be possible in the next iteration of some of the outer loops.
Figure~\ref{fig:refinepair} shows this schematically. 

The local search algorithm itself is basically the FM-algorithm
\cite{fiduccia1982lth}: For each
of the two blocks $A$, $B$ under consideration, a PE keeps a priority queue of nodes
eligible to move. The priority is based on the \emph{gain}, i.e., the decrease
in edge cut when the node is moved to the other side. Each node is moved at most
once within a single local search. The queues are initialized in random order
with the nodes at the partition boundary. We have tried several queue selection strategies:
\emph{Alternating} between $A$ and $B$ \cite{fiduccia1982lth},
\emph{MaxLoad} where always the heavier block gives a node, and \emph{TopGain},
where the queue promising larger gain is used. 
In order to achieve a good balance,
TopGain adopts the exception that MaxLoad is used when one of the blocks is overloaded. 
When not otherwise mentioned, we use TopGain with random tie breaking.
There is also a variant \emph{TopGainMaxLoad} that uses MaxLoad when
both queues promise the same gain.

The search is broken when
more than $\alpha\min\set{|A|,|B|}$ nodes  have been moved without yielding an improvement.
When the search stops, search is rolled back to the state with the lexicographically best
value of the tuple $(\Id{imbalance}, \Id{cutValue})$. Where \Id{imbalance} is
 $\max(0,\max(c(A)-L_{\max}, c(B)-L_{\max}))$.

\section{Experiments}\label{s:experiments}

\vspace{-4mm}
\paragraph*{Implementation.}
We have implemented the algorithm described above using C++ and MPI.  Overall,
our program consists of about 34\,000 lines of code.  Priority queues for the local
search are based on binary heaps. 
Hash tables use the library (extended STL) provided with the GCC compiler.

\vspace{-4mm}
\paragraph*{System.}
We have run our code on cluster with 200 nodes each equipped with two Quad-core
Intel Xeon processors (X5355) which run at a clock speed of 2.667 GHz, have 2x4
MB of level 2 cache each and run Suse Linux Enterprise 10 SP 1. All nodes are
attached to an InfiniBand 4X DDR interconnect which is characterized by its very
low latency of below 2 microseconds and a point to point bandwidth between two
nodes of more than 1300 MB/s. Our program was compiled using GCC 
Version 4.3.1 and optimization level 3 using OpenMPI 1.2.8.
Henceforth, a PE is one core of this machine.

\begin{table}[b]
\centering
\begin{tabular}{|l|r|r|}
\hline
graph & $n$ & $m$ \\
\hline
rgg17	 & $2^{17}$ 	& 1\,457\,506\\
rgg18 	 & $2^{18}$  	& 3\,094\,566\\
Delaunay17	 & $2^{17}$	& 786\,352\\
Delaunay18 	 & $2^{18}$ 	& 1\,572\,792\\
\hline
bcsstk29 	&	13\,992 &	605\,496\\
4elt 		&	15\,606 & 	91\,756\\
fesphere  	& 	16\,386 & 	98\,304\\
cti  		&	16\,840 & 	96\,464\\	
memplus 	&	17\,758 &	108\,384\\
cs4 	 	&       33\,499 & 	87\,716\\
pwt 		&	36\,519 & 	289\,588\\
bcsstk32 	&	44\,609 &	1\,970\,092\\	
body 		& 45\,087 	& 327\,468\\
t60k 		& 60\,005 	& 178\,880\\
wing  		& 62\,032 	& 243\,088\\
finan512 	& 74\,752 	& 522\,240\\	
ferotor 	& 99\,617  	& 1\,324\,862\\
\hline
bel & 463\,514 & 1\,183\,764\\
nld & 893\,041 & 2\,279\,080\\
\hline
af\_shell9 	& 504\,855 & 17\,084\,020\\
\hline
\end{tabular}
\hfill
\begin{tabular}{|l|r|r|}
\hline
graph & $n$ & $m$ \\
\hline
rgg20 		& $2^{20}$ 	& 13\,783\,240   \\
Delaunay20 		& $2^{20}$ 	& 12\,582\,744  \\
\hline
fetooth	 	& 78\,136  	& 905\,182\\
598a  		& 110\,971 	& 1\,483\,868\\
ocean 		& 143\,437 	& 819\,186\\
144  		& 144\,649 	& 2\,148\,786\\
wave 		& 156\,317 	& 2\,118\,662  \\
m14b 		& 214\,765 	& 3\,358\,036 \\
auto 		& 448\,695 	& 6\,629\,222 \\
\hline
deu & 4\,378\,446 & 10\,967\,174  \\
eur & 18\,029\,721 & 44\,435\,372 \\
\hline
af\_shell10 & 1\,508\,065 & 51\,164\,260  \\
\hline
coAuthorsDBLP 	& 299\,067 & 1\,955\,352\\       
citationCiteseer & 434\,102 & 32\,073\,440 \\       
\hline
\end{tabular}

\caption{
  Basic properties of the graphs from our 
  benchmark set. left: small to medium sized inputs, right: large instances.
  The latter class is split into five groups:
  geometric graphs, FEM graphs, street networks, sparse matrices, and social networks.
  Within their groups, the graphs are sorted by size.
}
\label{tab:instances}
\end{table}

\vspace{-4mm}
\paragraph*{Instances.}
We report experiments on two suites of instances summarized in
Table~\ref{tab:instances}. \Id{rggX} is a \emph{random geometric graph} with
$2^{X}$ nodes where nodes represent random points in the unit square and edges
connect nodes whose Euclidean distance is below $0.55 \sqrt{ \ln n / n }$.
This threshold was choosen in order to ensure that the graph is almost connected. 
\Id{DelaunayX} is the Delaunay triangulation of $2^{X}$
random points in the unit square.  Graphs \Id{bcsstk29}..\Id{fetooth} and
\Id{ferotor}..\Id{auto} come from Chris Walshaw's benchmark archive
\cite{walshaw2000mpm}.  Graphs \Id{bel}, \Id{nld}, \Id{deu} and \Id{eur} are
undirected versions of the road networks of Belgium, the Netherlands, Germany,
and Western Europe respectively, used in \cite{DSSW08}. Instances
\Id{af\_shell9} and \Id{af\_shell10} come from the Florida Sparse
Matrix Collection \cite{UFsparsematrixcollection}.  \Id{coAuthorsDBLP},
\Id{citationCiteseer} are examples of social networks taken from \cite{GSS08}.
Coordinate information is available for \Id{rggX}, \Id{DelaunayX}, the road
networks, \Id{bel}, \Id{nld}, \Id{deu} and \Id{eur}, and for the finite element
grahs \Id{feocean} and \Id{fetooth}.

%

For the number of partitions $k$ we choose the values
used in  \cite{walshaw2000mpm}: 2, 4, 8, 16, 32, 64.
Our default value for the allowed inbalance is 3 \% since this is one
of the values used in \cite{walshaw2000mpm} and the default value in Metis.

When not otherwise mentioned, we perform 10 repetitions of each run and report the
average result. When averaging over multiple instances, we use the geometric mean in order
to give every instance the same influence on the final figure.

\subsection{Configuring the Algorithm}\label{ss:parameters}

\newcommand{\onecol}[1]{\multicolumn{3}{c}{#1}}
\begin{table}[b]
\begin{center}
\begin{tabular}{l|ccc}
parameter & minimal & fast & strong \\\hline
rating    & \onecol{$\expansion^{*2}$}\\
matching  & \onecol{GPA}\\
stop contraction & \onecol{$n/60 k^2$}\\\hline
init. part. & \onecol{Scotch}\\
init. repeats & 1& 3 & 5 \\\hline
queue selection & \onecol{TopGain}\\
BFS search depth & 1 & 5 & 20\\
stop refinement & - & no change& $2\times$ no change \\ 
max. global iterations & 1 & 15 & 15 \\
local iterations & 1 & 3 & 5 \\
matching selection & \onecol{distr. edge coloring}\\
FM-patience $\alpha$ & 1 \% & 5 \% & 20 \%  \\\hline
avg. cut (geom.) & 2985 & 2910 & 2890 \\
avg. time (geom.)[s] & 0.67 & 1.29 & 2.10 \\
\end{tabular}
\end{center}
\caption{Parameter settings the for our main strategies.}
\label{tab:strategies}
\end{table}
Any multilevel algorithm has a considerable number of choices between
algorithmic components and tuning parameters. In the following we explore the
most important of these choices. In each case we will infer either a single
``good'' setting or two choices: the \Id{fast} setting aims at a low execution
time that still gives good partitioning quality and the \Id{strong}
setting targets good partitioning quality without investing an outrageous amount
of time. At no point we tune parameters specifically for one instance. All other
parameters are fixed at the default choices. When not otherwise mentioned, we
use the \Id{fast} parameter setting.
For some of the values we do not show experiments to save space and because the
experiments we did try do not give much new insight. Table~\ref{tab:strategies}
summarizes the settings. There is also a \Id{minimal} variant where for all 
parameters the smallest possible value is chosen. Although the minimal variant can be viewed
as overly crippled, it is useful when comparing to other, faster solvers.

\vspace{-4mm}
\paragraph*{Edge Ratings.} Table~\ref{tab:ratingMatching} shows the average performance
for different edge ratings.  Note that the plain edge weight is considerably worse
than the other ratings -- up to 8.8 \%.  The other ratings are fairly
close to each other and further experiments indicate that the remaining
differences heavily depend on the instances and other parameters of the
strategy. We adopt $\expansion^{*2}$ in the following.
\begin{table}[htb]
{\small\begin{center}
		\begin{tabular}{|l|r|r|r|r|}\hline
			Edge Rating & avg. & best. & avg. bal. & avg. t\\
			\hline 
			expansion$^*2$ & 2910 & 2819 & 1.025 & 1.29 \\
			expansion$^*$ & 2914 & 2815 & 1.025 & 1.30 \\
			innerOuter & 2914 & 2816 & 1.025 & 1.32 \\
			expansion & 2940 & 2841 & 1.025 & 1.31 \\
			weight & 3165 & 3010 & 1.026 & 1.40 \\
			\hline
		\end{tabular}
		\begin{tabular}{|l|r|r|r|r|}\hline
			Seq. Match. & avg. & best. & avg. bal. & avg. t\\
				\hline
			gpa & 2910 & 2819 & 1.025 & 1.29 \\
			shem & 2984 & 2883 & 1.025 & 1.29 \\
			greedy & 3854 & 3347 & 1.025 & 1.78 \\
		\hline
		\end{tabular}
		\caption{Results for KaPPa-Fast for different edge ratings and matching algorithms.}
\label{tab:ratingMatching}
\end{center}
}\end{table}

\vspace{-4mm}
\paragraph*{Sequential Matching Algorithm.} In Table~\ref{tab:ratingMatching}, we
see that the other algorithms have at least 2.5 \% worse edge cuts than GPA. 
Note that the
overall running time in both configurations is about the same -- although GPA is
slower than SHEM, this disadvantage is offset by less work in the refinement
phase. The greedy algorithm performs worse than the other strategies. This is
astonishing since in \cite{MauSan07} it produces fairly good results. Moreover,
in the sequential experiments in \cite{DAHoltgrewe} it also works well and
outperforms SHEM. Apparently, there are some negative interactions with the
parallelization here.

\vspace{-4mm}
\paragraph*{Initial Partitioning.} 
So far, we tried pMetis and Scotch for initial partitioning.
pMetis is about 4.7 \% worse than Scotch and only has slightly 
lower overall runtime. We therefore adopt it as our default initial partitioner.


\vspace{-4mm}
\paragraph*{Queue Selection.}
Table~\ref{tab:queueSelection} indicates that TopGain gives about 3.2 \% better solutions
than the more standard MaxLoad strategy. 
Interestingly, the details of the strategy are very important. Without resolving to MaxLoad in an overloaded situation we would not be able to fulfill the balance constraint. On the other hand,
even using MaxLoad for tie breaking we are already worse than the seemingly stupid
Alternating rule.

\begin{table}[b]
\makebox[1.1\textwidth][c]{\hspace*{-12mm}\small%
		\begin{tabular}{|l|r|r|r|r|}\hline
			Queue Sel. S. & avg. & best. & bal. & avg. t\\
			\hline
			TopGain & 2910 & 2819 & 1.025 & 1.29 \\
			Alternate & 2942 & 2839 & 1.024 & 1.27 \\
			TopgainMaxLoad & 2948 & 2855 & 1.014 & 1.22 \\
			MaxLoad & 3002 & 2899 & 1.005 & 1.34 \\
			\hline
		\end{tabular}%
		\begin{tabular}{|l|r|r|r|r|}\hline
			Variant & avg. & best. & bal. & avg. t\\
			KaPPa-Strong & 24227 & 23739 & 1.028 & 36.93 \\
			KaPPa-Fast & 24725 & 24254 & 1.028 & 21.40 \\
			KaPPa-Minimal & 26720 & 26005  & 1.028 & 5.94 \\
			seq. scotch & 26811 & - & 1.027 & 5.95 \\ 
			kmetis & 28705 & 26904 & 1.026 & 0.79 \\
			parmetis & 31523 & 30449 & 1.041 & 0.59 \\
			\hline

		\end{tabular}}
		\caption{Left: Results for KaPPa-Fast for different queue selection strategies. Right: Comparison with other tools.}
\label{tab:queueSelection}

\end{table}

\vspace{-4mm}
\paragraph*{Global Iterations, Local Iterations, BFS Depth, and Local Search Parameters.}
For these parameters we get the predictable effect that more work yields better
solutions albeit at a decreasing return on investment. It is then hard to say
what parameters would be optimal. Roughly, our fast strategy represents values
that yield execution times no more than 20 \% 
larger than for the smallest possible value. 
These increases in execution time add up to 63 \% more execution time than the fast strategy
on average.

\subsection{Comparison with other Partitioners}\label{ss:others}

We now switch to our suite of larger graphs since thats what \algname\ was
designed for and because we thus avoid the effect of overtuning our algorithm
parameters to the instances used for calibration.

Table~\ref{tab:queueSelection} compares the performances of \algname\ with Scotch, kMetis
(sequential) and parMetis (parallel).
Detailed, per instance results can be found in Appendix~\ref{a:detail}.
parMetis produces about 30 \% larger cuts than the strong variant of \algname,
27 \% more than the fast one, and still 18 \% more than the minimal one.
Note that this differences are much larger than what can be obtained by just
repeated runs, which gives only about 3 \% improvement for 10 repetitions. 
Moreover parMetis is not able to fully adhere to the balancing constraint.
On the other hand, parMetis is at least an order of magnitude faster. 

For kMetis the differences are 18 \%, 16 \% and 7\% respectively.
For Scotch, we get 10 \% for the strong variant, 8 \% for the fast variant,
and similar partitioning quality as for the weak variant.
Comparing average execution times of parallel \algname\ 
with the sequential algorithms scotch and kMetis makes little sense because
this depends a lot on the number of PEs used.

Although a large gap between the running times remains, the differences get
smaller if one only considers graphs for which the current implementation
of \algname\ was optimized: large graphs with coordinate information that
allows geometric prepartitioning. Table~\ref{tab:reallyLarge} in the appendix shows data for the
four graphs in our benchmark suite that have at least one million nodes and
coordinate information (\Id{rgg20}, \Id{Delaunay20}, \Id{deu}, \Id{eur}).
First note that for the European road network, \Id{eur}, \algname\ produces a several times
smaller cut than Metis. Apparently, Metis was not able at all to discover
the structure inherent in the network (e.g., due to waterbodies, mountains, and national borders).
\algname-minimal now outperforms Scotch, comes close to kMetis and is only 
a factor 3--6 slower than parMetis. Also note that the absolute execution times 
are in the range of a few seconds -- few applications working on such large
graphs will work on that time scale. Another interesting observation is that
none of the other algorithms consistently complies with the balance
constraint of 3 \%. This is astonishing since these graphs have a very ``harmless'' structure --
they are near planar (except for \Id{rgg}) and have low maximum degree).
It seems that our approach of careful, pairwise refinement successfully avoids such problems.


	

%
%

\begin{figure}[tb]
\begin{center}
\vspace*{-10mm}
\includegraphics[width=250pt]{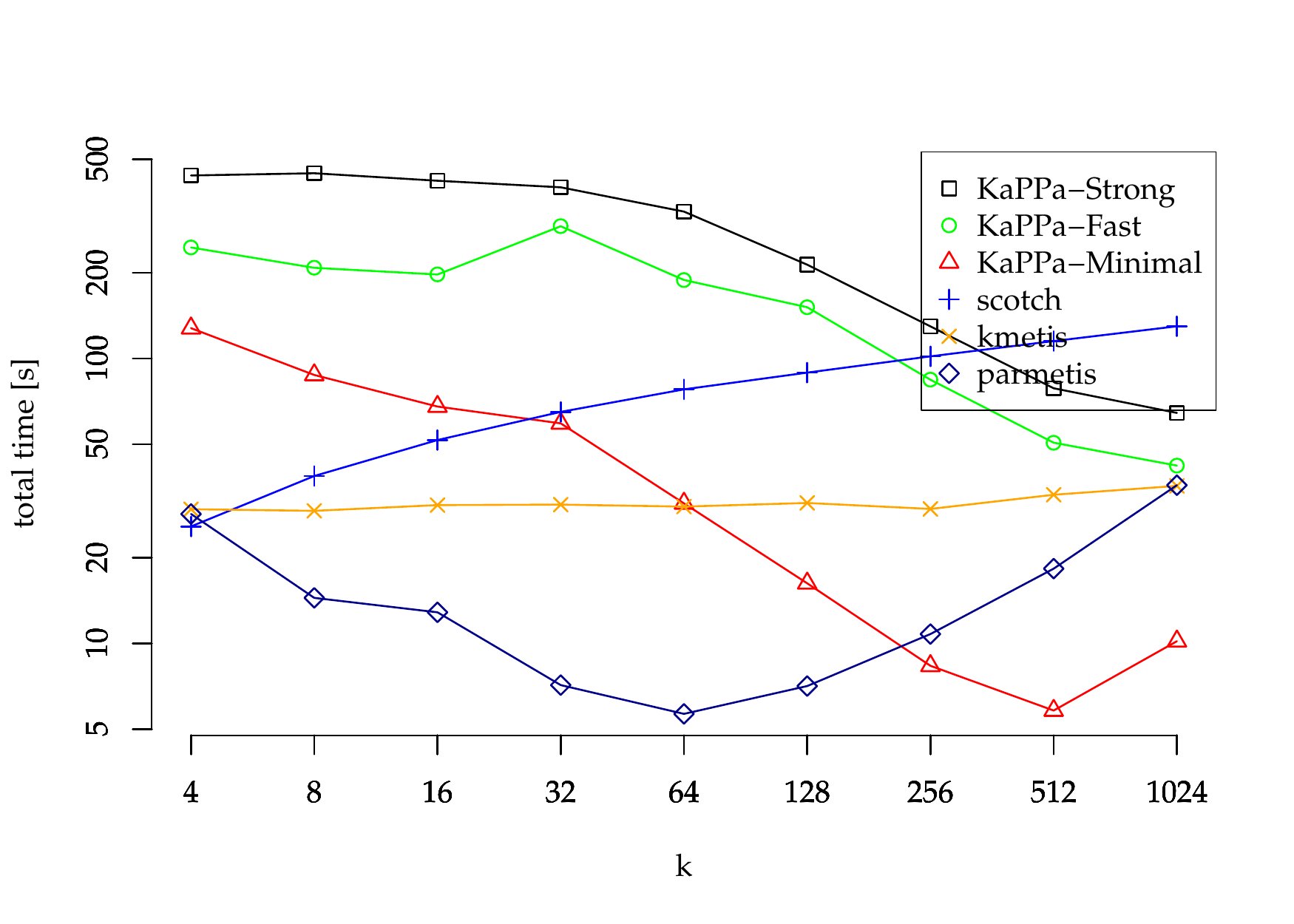}\\[-10mm]
\includegraphics[width=250pt]{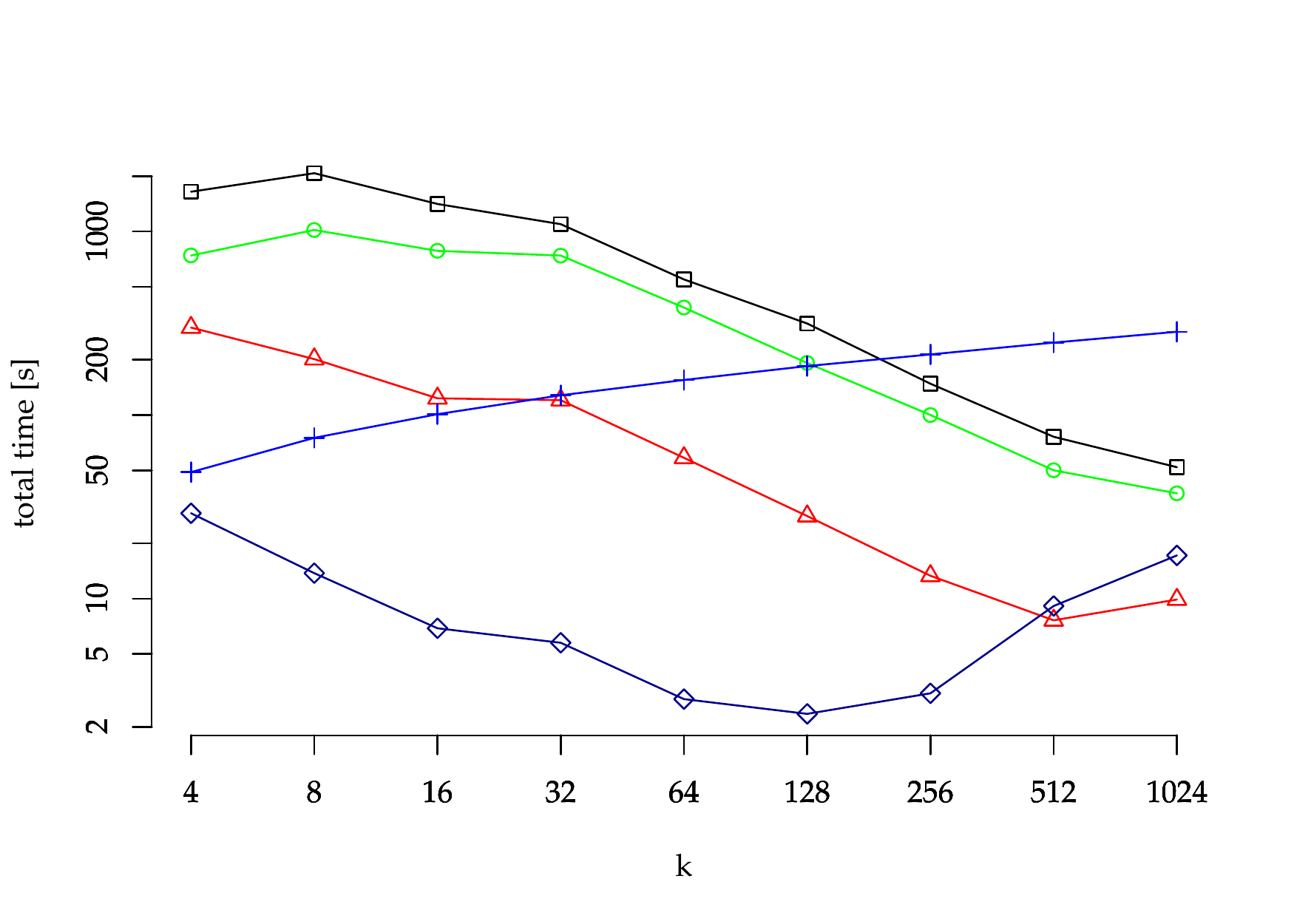}\\[-10mm]
\includegraphics[width=250pt]{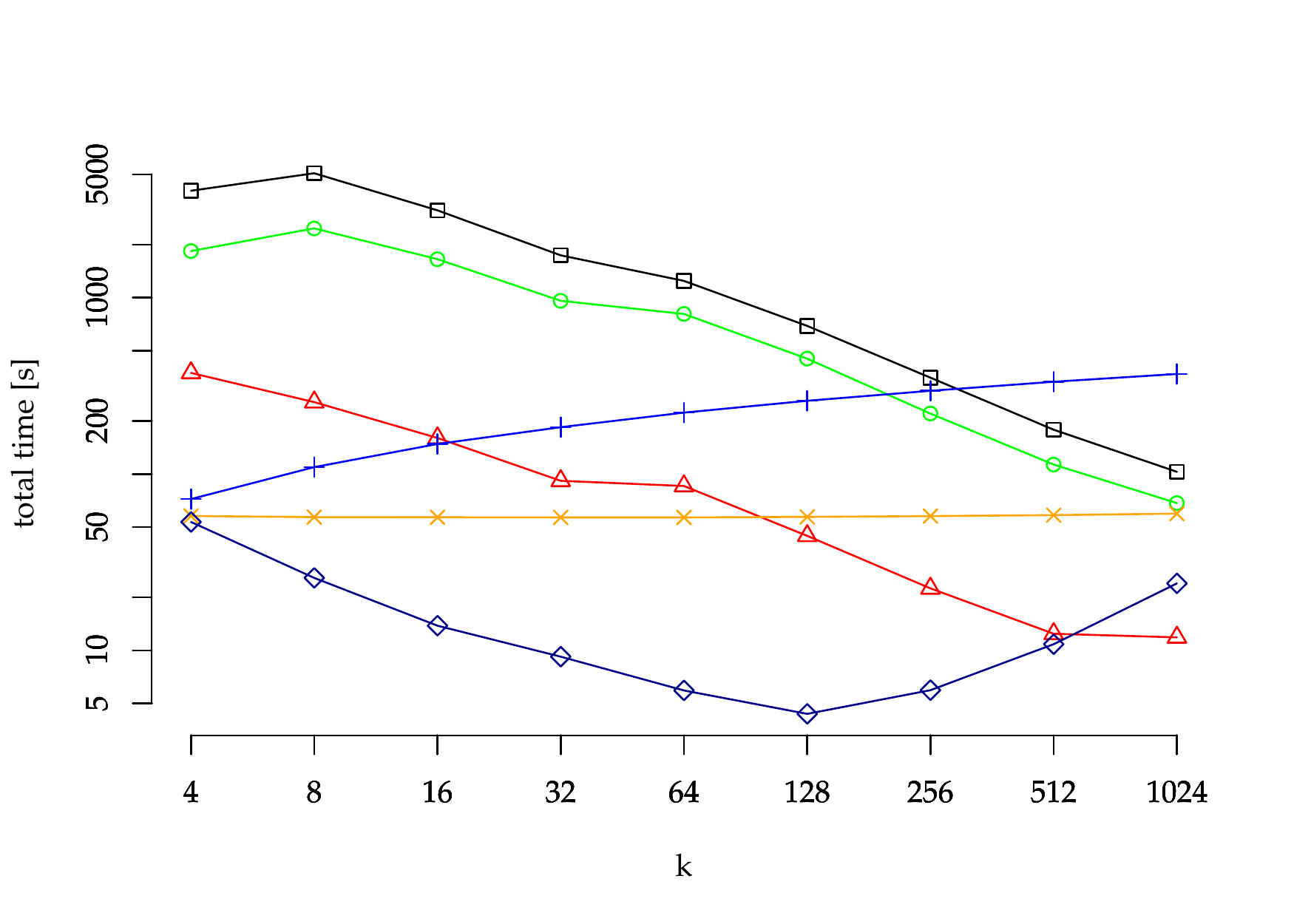}
\caption{Scalability for graphs eur, rgg25, and Delaunay25.}
\label{fig:scale}
\end{center}
\end{figure}

For the largest graphs available to us, we have scaled the number of processors further up to 1024. 
In Figure~\ref{fig:scale} we see that \algname\footnote{The minimal variant scales up to 512 PEs but this could be repaired by breaking the contraction later.} 
scales well all the way to the largest
number of processors, while parMetis reaches its limit of scalabilty at around 100 PEs.
Eventually, parMetis is slower than the fastest variant of \algname.

\subsection{The Walshaw Benchmark}\label{ss:benchmark}

We now apply \algname\ to Walshaw's benchmark archive
\cite{walshaw2000mpm,SWC04} using the rules used there, i.e., running time is no
issue but we want to achieve minimal cut values for $k\in \set{2, 4, 8, 16, 32,
  64}$ and balance parameter \frage{todo: balance
  0}$\epsilon\in\set{0.01,0.03,0.05}$. Thus, we further strengthen the strong
strategy: We try each of the edge ratings innerOuter, expansion$^*$, and
expansion$^{*2}$ 50 times; BFS search depth is 20; FM patience $\alpha=30$ \%.
Tables~\ref{tab:walshaw1}--\ref{tab:walshaw5} in the Appendix show the results
(left: \algname, right: best previous value) indicating an edge rating function
that achieved our result.  We obtain 54 improved entries for balance 5 \%, 46
improvements for 3 \%, and 31 improvements for balance 1 \%.  One interpretation
is that the improvement due to the TopGain queue selection strategy become less
effective for very small imbalance.  Indeed, for balance 0 TopGain yields no
improvements.\frage{todo: sth better ? then include table?}%
\footnote{However, the MaxLoad strategy given some slack on the balance
  constraint, yields good solutions that, for small $k$, are often fully
  balanced and yield improved values.}  For 11 out of 14 instances from the
large graphs we obtain improvements somewhere and for 9 out of 20 small
instances (for all but two of the small instances we sometimes find a solution
with the best known cut).  The biggest absolute improvement is observed for
instance \Id{add32} at 1 \% imbalance, and $k=64$ where the old partition cuts
45 \% more edges.  We obtain few improvements for $k=2$, perhaps still lacking
specialized techniques for that case. We have many improvements for $k=4$ going
down for smaller graphs and larger $k$. Perhaps this could be changed by
combining \algname\ with evolutionary techniques such as \cite{SWC04}. For large
$k$ we expect evolutionary methods to be superior to plain restarts that then
have trouble exploring a sufficient part of the solution space.













\section{Related Work}\label{s:related}

This paper is a summary and extension of the diploma theses
\cite{DASchulz,DAHoltgrewe}. There has been a huge amount of research on graph
partitioning so that we refer to overview papers such as
\cite{fjallstrom1998agp,SchKarKum00,Walshaw07} for a general overview. From
now on focus on issues closely related to the contributions of our paper. All
successful methods that are able to obtain good partitions for large real world
graphs are based on the multilevel principle outlined in
Section~\ref{s:preliminaries}. The basic idea can be traced back to multigrid
solvers for solving systems of linear equations \cite{Sou35,Fedorenko61} but
more recent practical methods are based on mostly graph theoretic aspects in
particular edge contraction and local search.  Well known software packages
based on this approach include Chaco \cite{Chaco}, Jostle \cite{Walshaw07},
Metis \cite{SchKarKum00}, Party \cite{Party}, and Scotch \cite{Scotch}.  
While Chaco and Party are no longer developed and have no parallel version,
the others have been parallelized also.
Probably the
fastest available parallel code is the parallel version of Metis, parMetis.
However, its partitioning quality is worse than the sequential version kMetis.
In general it seems to be the case that previous parallelizations came with a
penalty in partitioning quality.  In contrast, our parallelization approach seems to
\emph{improve} partitioning quality.

The parallel version of Jostle \cite{Walshaw07} is similar to our approach since
it applies local search to pairs of neighboring partitions. However, this
parallelization has problems maintaining the balance of the partitions since at
any particular time, it is difficult to say how many nodes are assigned to a
particular block. We solve this problems by performing concurrent local searches
only on independent pairs of partitions. 

PT-Scotch, the parallel version of Scotch is based on
recursive bipartitioning. This is more difficult to parallelize than direct
$k$-partitioning since in the initial bipartition, there is less parallelism
available.  The unused processor power is used by performing several independent
attempts in parallel. The involved communication effort is reduced by considering only nodes
close to boundary of the current partitioning (band-refinement). We also use band-refinement
but using a different algorithm and with much less replication of work.

DiBaP \cite{meyerhenke2008ndb} is a multi-level graph partitioning package
based on diffusion.  It currently yields the best partitioning results for the
biggest graphs in \cite{Walshawbench} but has no scalable
parallelization.

Most previous approaches use the edge weight to quantify with which preference
it is included into a matching.
In \cite{AboKar06}, many different edge
ratings are considered. However all of them use a very simple rating as the
primary sorting criterion. In contrast, our approach genuinely combines the two
sometimes conflicting criteria of contracting heavy edges and light vertices.

The need for fast, (near) linear time algorithms for approximate weighted
matchings in hierarchical graph partitioning has been a major motivation for
developing such algorithms \cite{Preis99,DH03a,DH03c,PS04,MauSan07}. In contrast
to the heavy edge matching algorithms used in most systems, these schemes give
approximation guarantees of 1/2 \cite{Preis99,DH03a} or 2/3 \cite{DH03c,PS04}.
In \cite{MauSan07} we developed another 1/2 algorithm that turned out to be even
better than the 2/3 algorithms in many practical cases. Interestingly, only few
of these results have so far found their way into actual graph
partitioners. One contribution of our paper is to try them out.

\section{Conclusions and Future Work}\label{s:conclusions}

We have demonstrated that high quality graph partitioning can be done in
parallel in a scalable way. This success is due to several
innovations/observations that might also work in the framework of other graph
partitioning and graph clustering systems: Edge rating functions that take into
account other aspects than edge weight give considerably better results (8.8 \%
on the average for the experiments in Section~\ref{ss:parameters}). In
particular, it seems that discouraging heavy nodes leads to much more uniform
contraction all over the graph. High quality matching algorithms like GPA also
yield a few percent improvement. In particular, the computational overhead for
these algorithms is not affecting the overall runtime of a high quality graph
partitioner, presumably because of less work in the refinement phase.  FM-style
local search can also yield improved quality if the highest gain queue
is selected if possible. Feasibility can be maintained using an exception for
overloaded blocks. Again, a few percent improvement in solution quality
can be obtained.  Perhaps the most surprising result is that localizing the
local search to two blocks at a time does at the same time enable parallelization
and \emph{improve} partitioning quality compared to global local search.
Although the individual improvement due to each improvement is relatively small,
they add up to a sizable overall improvement. Also note that within a less tuned system,
adding one of the improvements may have a larger effect than in a code with
all improvments at once.

The current implementation of \algname\ is a research prototype rather than
a widely usable tool. 
But considering its good results, we want to further improve it
and advance it into a fully usable system usable for all kinds of inputs 
ranging from small graphs better handled by a lean sequential implementation
to huge graphs with billions of nodes.

Besides many implementation issues that will hopefully improve
execution time, the main conceptual task will be a generalization of the
interface. We want a system where the number of partitioning PEs $P$ and the
number of blocks $k$ can be chosen independently. This is rather straight
forward when $k>P$ since this actually increases the amount of parallelism. For
$k<P$, we could simply assign more PEs to the same local search (using different
seeds).  This would improve quality but reduces scalability and will not work
for huge graphs where block sizes may exceed local memory size. Therefore, we
need a parallel refinement algorithm working on only two neighboring blocks. We
also want to improve performance for graphs that are neither prepartitioned nor
equipped with coordinates. The easiest solution for moderate $P$ will be to use
parMetis for initial partitioning. For very large systems we want to develop a
very fast prepartitioner that works purely graph theoretically. A core component
will be fast scalable parallel contraction. There will also be further issues
when \algname\ is generalized for graph clustering, hypergraph partitioning, or
repartitioning. Besides improving the functionality of \algname, there are also many
ways to improve its basic performance. In particular, it would be desirable to
implement a more efficient representation of the distributed graph data
structure.

Besides improving functionality of \algname, many interesting research questions
remain. For example, one should investigate rating functions for edge contraction more
systematically. Other refinement algorithms, e.g., based on flows or diffusion could be
tried within our framework of pairwise refinement. 

{\small \bibliographystyle{plain}
\bibliography{paper}}
\clearpage
\begin{appendix}

\section{Detailed Results for the Large Instances.}\label{a:detail}

\begin{table}[htb]
	\begin{center}
		\begin{tabular}{|l|r|l|r|r|r|r|r|}\hline

			alg. & $k$ & graph &  avg. cut & best. cut. & avg. balance & avg. runtime\\
			\hline 
			\algname-strong&64&rgg20 & 35354 &  34778 & 1.030 & 11.62 \\
			\algname-strong&64&Delaunay20 & 25179 &  24799 & 1.030 & 22.04 \\
			\algname-strong&64&deu & 4093 &  4021 & 1.029 & 49.55 \\
			\algname-strong&64&eur & 5393 &  5290 & 1.030 & 308.17 \\
			\hline 
			\algname-fast&64&rgg20 & 35539 &  35086 & 1.030 & 9.95 \\
			\algname-fast&64&Delaunay20 & 25129 &  24946 & 1.030 & 12.83 \\
			\algname-fast&64&deu & 4146 &  4078 & 1.029 & 31.63 \\
			\algname-fast&64&eur & 5538 &  5448 & 1.030 & 183.98 \\
\hline
			\algname-minimal&64&rgg20 & 35629 &  35252 & 1.030 & 2.09 \\
			\algname-minimal&64&Delaunay20 & 27001 &  26314 & 1.029 & 1.79 \\
			\algname-minimal&64&deu & 4317 &  4193 & 1.029 & 5.97 \\
			\algname-minimal&64&eur & 5770 &  5569 & 1.029 & 29.64 \\
\hline
			Scotch &  64 & rgg20 & 38815 &  38815 & 1.031 & 9.84 \\
			Scotch &  64 & Delaunay20 & 26163 &  26163 & 1.037 & 7.36 \\
			Scotch &  64 & deu & 4978 &  4978 & 1.028 & 19.52 \\
			Scotch &  64 & eur & 6772 &  6772 & 1.031 & 77.41 \\
\hline 
			kMetis&64&rgg20 & 42465 &  41066 & 1.030 & 1.58 \\
			kMetis&64&Delaunay20 & 28543 &  28318 & 1.030 & 1.21 \\
			kMetis&64&deu & 5385 &  5147 & 1.029 & 5.31 \\
			kMetis&64&eur & 12738 &  11313 & 1.070 & 30.30 \\

			\hline 
			parMetis&64&rgg20 & 43545 &  42863 & 1.050 & 0.55 \\
			parMetis&64&Delaunay20 & 30321 &  29535 & 1.047 & 0.65 \\
			parMetis&64&deu & 7273 &  7083 & 1.027 & 0.91 \\
			parMetis&64&eur & 16427 &  14976 & 1.025 & 5.65 \\
			\hline

		\end{tabular}
	 \end{center}  
	 \caption{Performance for the  largest graphs with coordinate information.}
\label{tab:reallyLarge}
\end{table}

\begin{table}[h]	
 \begin{center}
		\begin{tabular}{|l|r|r|r|r|r|}\hline

			graph &  avg. cut & best. cut. & avg. balance & avg. runtime\\

			\hline 
			rgg20 & 15442 &  15039 & 1.029 & 7.20 \\
			Delaunay20 & 11533 &  11307 & 1.028 & 6.31 \\
			fetooth & 19813 &  19559 & 1.029 & 0.65 \\
			598a & 28596 &  27983 & 1.030 & 6.76 \\
			feocean & 9553 &  9457 & 1.029 & 0.70 \\
			144 & 41977 &  40264 & 1.030 & 6.63 \\
			wave & 47270 &  46293 & 1.029 & 1.47 \\
			m14b & 49397 &  48769 & 1.030 & 6.41 \\
			auto & 86001 &  84236 & 1.030 & 12.25 \\
			deu & 1656 &  1593 & 1.029 & 21.79 \\
			eur & 2048 &  1931 & 1.026 & 94.56 \\
			afshell10 & 175918 &  174677 & 1.029 & 17.00 \\
			coAuthorsDBLP & 163463 &  161842 & 1.030 & 9.99 \\
			citationCiteseer & 254914 &  253359 & 1.030 & 20.07 \\
			\hline
		
		\end{tabular}
	 \end{center}  
	 \caption{KaPPa-Minimal $k=16$}
\end{table}
\begin{table}[H]	
	\begin{center}
		\begin{tabular}{|l|r|r|r|r|r|}\hline

			graph &  avg. cut & best. cut. & avg. balance & avg. runtime\\

			\hline 
			rgg20 & 24164 &  23842 & 1.029 & 3.94 \\
			Delaunay20 & 18179 &  17993 & 1.029 & 3.33 \\
			fetooth & 28391 &  28070 & 1.030 & 0.53 \\
			598a & 43741 &  43111 & 1.030 & 7.74 \\
			feocean & 15657 &  15465 & 1.030 & 0.47 \\
			144 & 62171 &  61774 & 1.030 & 8.79 \\
			wave & 68620 &  68085 & 1.030 & 1.02 \\
			m14b & 73598 &  72484 & 1.030 & 8.12 \\
			auto & 133723 &  131545 & 1.030 & 20.23 \\
			deu & 2711 &  2626 & 1.029 & 11.50 \\
			eur & 3386 &  3202 & 1.029 & 55.63 \\
			afshell10 & 275149 &  270249 & 1.030 & 9.25 \\
			coAuthorsDBLP & 172830 &  171784 & 1.030 & 8.57 \\
			citationCiteseer & 285710 &  278587 & 1.030 & 19.83 \\
			\hline

		\end{tabular}
	 \end{center}  
	 \caption{KaPPa-Minimal $k=32$}
\end{table}
\begin{table}[H]
	\begin{center}
		\begin{tabular}{|l|r|r|r|r|r|}\hline

			graph &  avg. cut & best. cut. & avg. balance & avg. runtime\\

			\hline 
			rgg20 & 35629 &  35252 & 1.030 & 2.09 \\
			Delaunay20 & 27001 &  26314 & 1.029 & 1.79 \\
			fetooth & 39095 &  38423 & 1.029 & 0.62 \\
			598a & 61924 &  61396 & 1.029 & 6.21 \\
			feocean & 24275 &  24147 & 1.030 & 0.51 \\
			144 & 86950 &  86067 & 1.030 & 8.16 \\
			wave & 93424 &  92366 & 1.030 & 1.03 \\
			m14b & 107173 &  106361 & 1.030 & 10.24 \\
			auto & 187424 &  185836 & 1.030 & 25.39 \\
			deu & 4317 &  4193 & 1.029 & 5.97 \\
			eur & 5770 &  5569 & 1.029 & 29.64 \\
			afshell10 & 404085 &  400378 & 1.030 & 4.82 \\
			coAuthorsDBLP & 180724 &  180059 & 1.030 & 15.82 \\
			citationCiteseer & 315062 &  313465 & 1.030 & 22.86 \\
			\hline

		\end{tabular}
	 \end{center}  

	 \caption{KaPPa-Minimal $k=64$}
\end{table}
\begin{table}[H]	
 \begin{center}
		\begin{tabular}{|l|r|r|r|r|r|}\hline

			graph &  avg. cut & best. cut. & avg. balance & avg. runtime\\

			\hline 
			rgg20 & 15339 &  15013 & 1.029 & 24.61 \\
			Delaunay20 & 11061 &  10882 & 1.029 & 48.05 \\
			fetooth & 18524 &  18198 & 1.030 & 3.55 \\
			598a & 26887 &  26670 & 1.030 & 12.51 \\
			feocean & 8469 &  8294 & 1.030 & 3.04 \\
			144 & 39492 &  39266 & 1.030 & 17.53 \\
			wave & 45202 &  44936 & 1.030 & 10.73 \\
			m14b & 46108 &  45931 & 1.030 & 19.27 \\
			auto & 80683 &  79711 & 1.030 & 58.20 \\
			deu & 1618 &  1556 & 1.027 & 78.82 \\
			eur & 1935 &  1907 & 1.028 & 295.81 \\
			afshell10 & 166480 &  165625 & 1.030 & 69.97 \\
			coAuthorsDBLP & 150272 &  149302 & 1.030 & 66.47 \\
			citationCiteseer & 203302 &  198450 & 1.030 & 85.02 \\
			\hline

		\end{tabular}
	 \end{center}  
	 \caption{KaPPa-Fast $k=16$}
\end{table}
\begin{table}[H]	
	\begin{center}
		\begin{tabular}{|l|r|r|r|r|r|}\hline

			graph &  avg. cut & best. cut. & avg. balance & avg. runtime\\

			\hline 
			rgg20 & 24222 &  23383 & 1.030 & 16.93 \\
			Delaunay20 & 17150 &  16814 & 1.030 & 24.44 \\
			fetooth & 26677 &  26404 & 1.030 & 2.92 \\
			598a & 41186 &  40928 & 1.030 & 11.91 \\
			feocean & 14042 &  13618 & 1.030 & 2.15 \\
			144 & 58652 &  58175 & 1.030 & 16.03 \\
			wave & 64532 &  64004 & 1.030 & 8.19 \\
			m14b & 69223 &  68715 & 1.030 & 17.99 \\
			auto & 125876 &  124920 & 1.030 & 46.44 \\
			deu & 2641 &  2535 & 1.029 & 41.93 \\
			eur & 3314 &  3231 & 1.030 & 306.52 \\
			afshell10 & 255746 &  252487 & 1.030 & 52.00 \\
			coAuthorsDBLP & 163767 &  162577 & 1.030 & 58.93 \\
			citationCiteseer & 233459 &  229629 & 1.030 & 83.14 \\
			\hline

		\end{tabular}
	 \end{center}  
	 \caption{KaPPa-Fast $k=32$}
\end{table}
\begin{table}[H]
	\begin{center}
		\begin{tabular}{|l|r|r|r|r|r|}\hline

			graph &  avg. cut & best. cut. & avg. balance & avg. runtime\\

			\hline 
			rgg20 & 35539 &  35086 & 1.030 & 9.95 \\
			Delaunay20 & 25129 &  24946 & 1.030 & 12.83 \\
			fetooth & 36992 &  36795 & 1.029 & 2.57 \\
			598a & 59233 &  59026 & 1.029 & 9.64 \\
			feocean & 21973 &  21809 & 1.030 & 2.02 \\
			144 & 82493 &  82029 & 1.030 & 12.05 \\
			wave & 89297 &  88924 & 1.030 & 6.09 \\
			m14b & 101861 &  101410 & 1.030 & 17.46 \\
			auto & 178119 &  177461 & 1.030 & 44.14 \\
			deu & 4146 &  4078 & 1.029 & 31.63 \\
			eur & 5538 &  5448 & 1.030 & 183.98 \\
			afshell10 & 384140 &  380225 & 1.030 & 29.43 \\
			coAuthorsDBLP & 174411 &  173629 & 1.030 & 65.76 \\
			citationCiteseer & 269854 &  268188 & 1.030 & 86.06 \\
			\hline

		\end{tabular}
	 \end{center}  

	 \caption{KaPPa-Fast $k=64$}
\end{table}
\begin{table}[H]	
 \begin{center}
		\begin{tabular}{|l|r|r|r|r|r|}\hline

			graph &  avg. cut & best. cut. & avg. balance & avg. runtime\\

			\hline 
			rgg20 & 15199 &  14953 & 1.029 & 35.86 \\
			Delaunay20 & 11008 &  10816 & 1.027 & 67.92 \\
			fetooth & 18570 &  18302 & 1.030 & 7.18 \\
			598a & 26825 &  26467 & 1.030 & 17.74 \\
			feocean & 8350 &  8188 & 1.030 & 5.62 \\
			144 & 39319 &  39010 & 1.030 & 26.04 \\
			wave & 45048 &  44831 & 1.030 & 20.54 \\
			m14b & 45762 &  45352 & 1.030 & 28.11 \\
			auto & 79769 &  78713 & 1.030 & 87.41 \\
			deu & 1616 &  1550 & 1.027 & 105.96 \\
			eur & 1900 &  1760 & 1.027 & 497.93 \\
			afshell10 & 166427 &  165025 & 1.030 & 106.63 \\
			coAuthorsDBLP & 145975 &  145031 & 1.030 & 105.61 \\
			citationCiteseer & 176690 &  171233 & 1.030 & 142.01 \\
			\hline

		\end{tabular}
	 \end{center}  
	 \caption{KaPPa-Strong $k=16$}
\end{table}
\begin{table}[H]	
	\begin{center}
		\begin{tabular}{|l|r|r|r|r|r|}\hline

			graph &  avg. cut & best. cut. & avg. balance & avg. runtime\\

			\hline 
			rgg20 & 23917 &  23430 & 1.029 & 26.04 \\
			Delaunay20 & 17086 &  16813 & 1.030 & 42.67 \\
			fetooth & 26617 &  26397 & 1.030 & 5.28 \\
			598a & 41190 &  40946 & 1.030 & 18.16 \\
			feocean & 13815 &  13593 & 1.030 & 4.34 \\
			144 & 58631 &  58331 & 1.030 & 24.60 \\
			wave & 64390 &  63981 & 1.030 & 14.94 \\
			m14b & 69075 &  68107 & 1.030 & 29.94 \\
			auto & 125500 &  124606 & 1.030 & 71.77 \\
			deu & 2615 &  2548 & 1.029 & 73.17 \\
			eur & 3291 &  3186 & 1.029 & 417.52 \\
			afshell10 & 255535 &  253525 & 1.030 & 80.85 \\
			coAuthorsDBLP & 161073 &  160225 & 1.030 & 106.63 \\
			citationCiteseer & 207559 &  203989 & 1.030 & 140.53 \\
			\hline

		\end{tabular}
	 \end{center}  
	 \caption{KaPPa-Strong $k=32$}
\end{table}
\begin{table}[H]
	\begin{center}
		\begin{tabular}{|l|r|r|r|r|r|}\hline

			graph &  avg. cut & best. cut. & avg. balance & avg. runtime\\

			\hline 
			rgg20 & 35354 &  34778 & 1.030 & 11.62 \\
			Delaunay20 & 25179 &  24799 & 1.030 & 22.04 \\
			fetooth & 37002 &  36862 & 1.029 & 4.71 \\
			598a & 59387 &  59148 & 1.029 & 14.15 \\
			feocean & 21859 &  21636 & 1.030 & 3.68 \\
			144 & 82452 &  82286 & 1.030 & 19.11 \\
			wave & 88964 &  88376 & 1.030 & 12.51 \\
			m14b & 101455 &  101053 & 1.030 & 25.26 \\
			auto & 177595 &  177038 & 1.030 & 62.64 \\
			deu & 4093 &  4021 & 1.029 & 49.55 \\
			eur & 5393 &  5290 & 1.030 & 308.17 \\
			afshell10 & 382923 &  379125 & 1.030 & 43.01 \\
			coAuthorsDBLP & 172132 &  171194 & 1.030 & 111.90 \\
			citationCiteseer & 249544 &  246150 & 1.030 & 146.65 \\
			\hline

		\end{tabular}
	 \end{center}  

	 \caption{KaPPa-Strong $k=64$}
\end{table}

\begin{table}[H]
	\begin{center}
		\begin{tabular}{|l|r|r|r|r|r|}\hline

			graph &  avg. cut & best. cut. & avg. balance & avg. runtime\\

			\hline 
			rgg20 & 18125 &  17498 & 1.021 & 1.53 \\
			Delaunay20 & 12440 &  11854 & 1.016 & 1.14 \\
			fetooth & 20386 &  20035 & 1.029 & 0.09 \\
			598a & 28854 &  27857 & 1.030 & 0.17 \\
			feocean & 10377 &  10115 & 1.029 & 0.13 \\
			144 & 43041 &  42861 & 1.030 & 0.24 \\
			wave & 49000 &  48404 & 1.030 & 0.22 \\
			m14b & 49269 &  48314 & 1.029 & 0.36 \\
			auto & 89139 &  85562 & 1.030 & 0.91 \\
			deu & 2161 &  2041 & 1.007 & 5.19 \\
			eur & 9395 &  3519 & 1.030 & 30.58 \\
			afshell10 & 188765 &  184350 & 1.014 & 3.06 \\
			coAuthorsDBLP & 139658 &  138334 & 1.031 & 0.98 \\
			citationCiteseer & 157011 &  153588 & 1.031 & 1.05 \\
			\hline

		\end{tabular}
	 \end{center}  

	 \caption{KMetis $k=16$}
\end{table}
\begin{table}[H]
	\begin{center}
		\begin{tabular}{|l|r|r|r|r|r|}\hline

			graph &  avg. cut & best. cut. & avg. balance & avg. runtime\\

			\hline 
			rgg20 & 18760 &  18193 & 1.048 & 0.39 \\
			Delaunay20 & 13126 &  12806 & 1.043 & 0.35 \\
			fetooth & 20686 &  20255 & 1.046 & 0.06 \\
			598a & 29858 &  29308 & 1.047 & 0.17 \\
			feocean & 10212 &  9951 & 1.043 & 0.06 \\
			144 & 43019 &  41841 & 1.050 & 0.19 \\
			wave & 49981 &  49537 & 1.048 & 0.09 \\
			m14b & 49621 &  47697 & 1.048 & 0.28 \\
			auto & 87057 &  84900 & 1.047 & 0.54 \\
			deu & 3166 &  3063 & 1.009 & 1.62 \\
			eur & 6861 &  5576 & 1.073 & 12.85 \\
			afshell10 & 191995 &  189925 & 1.048 & 0.74 \\
			coAuthorsDBLP & 193580 &  190892 & 1.044 & 1.44 \\
			citationCiteseer & 197095 &  197095 & 1.047 & 1.41 \\
			\hline

		\end{tabular}
	 \end{center}  
	 \caption{parMetis $k=16$}
\end{table}
\begin{table}[H]
	\begin{center}
		\begin{tabular}{|l|r|r|r|r|r|}\hline

			graph &  avg. cut & best. cut. & avg. balance & avg. runtime\\

			\hline 
			rgg20 & 28495 &  27765 & 1.029 & 1.58 \\
			Delaunay20 & 19304 &  18816 & 1.029 & 1.18 \\
			fetooth & 29052 &  28547 & 1.030 & 0.10 \\
			598a & 44213 &  43256 & 1.030 & 0.19 \\
			feocean & 16877 &  16565 & 1.030 & 0.15 \\
			144 & 62481 &  61716 & 1.030 & 0.26 \\
			wave & 68604 &  68062 & 1.030 & 0.25 \\
			m14b & 74135 &  72746 & 1.030 & 0.40 \\
			auto & 134086 &  133026 & 1.030 & 0.99 \\
			deu & 3445 &  3319 & 1.019 & 5.28 \\
			eur & 9442 &  7424 & 1.078 & 30.81 \\
			afshell10 & 291590 &  289400 & 1.027 & 3.13 \\
			coAuthorsDBLP & 160373 &  159032 & 1.030 & 1.19 \\
			citationCiteseer & 201073 &  197839 & 1.031 & 1.19 \\
			\hline

		\end{tabular}
	 \end{center}  

	 \caption{KMetis $k=32$}
\end{table}
\begin{table}[H]
	\begin{center}
		\begin{tabular}{|l|r|r|r|r|r|}\hline

			graph &  avg. cut & best. cut. & avg. balance & avg. runtime\\

			\hline 
			rgg20 & 29227 &  28650 & 1.049 & 0.22 \\
			Delaunay20 & 20141 &  19803 & 1.045 & 0.21 \\
			fetooth & 28790 &  28513 & 1.043 & 0.07 \\
			598a & 44422 &  43968 & 1.046 & 0.49 \\
			feocean & 16259 &  16010 & 1.040 & 0.05 \\
			144 & 62673 &  62244 & 1.049 & 0.51 \\
			wave & 70365 &  70072 & 1.048 & 0.15 \\
			m14b & 76447 &  75356 & 1.049 & 0.52 \\
			auto & 137913 &  137047 & 1.047 & 0.70 \\
			deu & 4858 &  4703 & 1.034 & 0.87 \\
			eur & 9616 &  8366 & 1.072 & 7.22 \\
			afshell10 & 293110 &  289275 & 1.048 & 0.35 \\
			coAuthorsDBLP & 211756 &  209846 & 1.046 & 1.59 \\
			citationCiteseer & 212524 &  212524 & 1.050 & 1.56 \\
			\hline

		\end{tabular}
	 \end{center}  

	 \caption{parMetis $k=32$}
\end{table}

\begin{table}[H]
	\begin{center}
		\begin{tabular}{|l|r|r|r|r|r|}\hline

			graph &  avg. cut & best. cut. & avg. balance & avg. runtime\\

			\hline 
			rgg20 & 42465 &  41066 & 1.030 & 1.58 \\
			Delaunay20 & 28543 &  28318 & 1.030 & 1.21 \\
			fetooth & 39381 &  39233 & 1.030 & 0.12 \\
			598a & 62703 &  61888 & 1.030 & 0.22 \\
			feocean & 24531 &  24198 & 1.030 & 0.17 \\
			144 & 87208 &  86534 & 1.030 & 0.30 \\
			wave & 94083 &  92148 & 1.030 & 0.29 \\
			m14b & 108141 &  107384 & 1.031 & 0.44 \\
			auto & 189699 &  188555 & 1.030 & 1.08 \\
			deu & 5385 &  5147 & 1.029 & 5.31 \\
			eur & 12738 &  11313 & 1.070 & 30.30 \\
			afshell10 & 427047 &  421285 & 1.030 & 3.18 \\
			coAuthorsDBLP & 176485 &  174402 & 1.033 & 1.42 \\
			citationCiteseer & 244330 &  242677 & 1.033 & 1.41 \\
			\hline

		\end{tabular}
	 \end{center}  

	 \caption{KMetis $k=64$}
\end{table}
\begin{table}[H]
	\begin{center}
		\begin{tabular}{|l|r|r|r|r|r|}\hline

			graph &  avg. cut & best. cut. & avg. balance & avg. runtime\\

			\hline 
			rgg20 & 43545 &  42863 & 1.050 & 0.55 \\
			Delaunay20 & 30321 &  29535 & 1.047 & 0.65 \\
			fetooth & 39477 &  38790 & 1.047 & 0.56 \\
			598a & 63688 &  62936 & 1.047 & 1.82 \\
			feocean & 26249 &  25912 & 1.039 & 0.12 \\
			144 & 87967 &  87163 & 1.047 & 1.58 \\
			wave & 95758 &  94605 & 1.049 & 0.44 \\
			m14b & 108546 &  107125 & 1.049 & 1.98 \\
			auto & 194958 &  192198 & 1.047 & 1.69 \\
			deu & 7273 &  7083 & 1.027 & 0.91 \\
			eur & 16427 &  14976 & 1.025 & 5.65 \\
			afshell10 & 435995 &  433525 & 1.049 & 0.20 \\
			coAuthorsDBLP & 218798 &  217403 & 1.050 & 2.32 \\
			citationCiteseer & 219850 &  219850 & 1.046 & 2.32 \\
			\hline

		\end{tabular}
	 \end{center}  
	 \caption{parMetis $k=64$}
\end{table}


\begin{landscape}

\begin{table}[p]
\begin{center}{\small
\begin{tabular}{|l|r|r|r|r|r|r|r|r|r|r|r|r|}\hline
 Graph  & \multicolumn{2}{|c|}{2} & \multicolumn{2}{|c|}{4} & \multicolumn{2}{|c|}{8} & \multicolumn{2}{|c|}{16} & \multicolumn{2}{|c|}{32} & \multicolumn{2}{|c|}{64}\\
\hline	
	3elt			&		**	90			&			\textbf{89}		&		+	201				&			\textbf{199}	&		*	354				&			\textbf{342}	&	*		597				&			\textbf{569}		&		*	1008			&			\textbf{969}		&		*	1629			&			\textbf{1564}	\\	
	add20			&		*	618			&			\textbf{594}	&		*	1190			&			\textbf{1177}	&		*	1752			&			\textbf{1704}	&	+		2141			&			\textbf{2121}		&		*	\textbf{2594}	&			2687				&		*	\textbf{3082}	&			3236			\\	
	data			&		+	191			&			\textbf{188}	&		*	\textbf{383}	&			383				&		*	664				&			\textbf{660}	&	**		1169			&			\textbf{1162}		&		*	1912			&			\textbf{1865}		&		*	2949			&			\textbf{2885}		\\	
	uk				&		*	20			&			\textbf{19}		&		+	44				&			\textbf{42}		&		*	88				&			\textbf{84}		&	+		159				&			\textbf{152}		&		*	273				&			\textbf{258}		&		**	445				&			\textbf{438}		\\	
	add32			&		**	\textbf{10}	&			10				&		**	\textbf{33}		&			33				&		**	\textbf{66}		&			\textbf{66}		&	+		124				&			\textbf{117}		&		+	223				&			\textbf{212}		&		*	\textbf{495}	&			720					\\	
	bcsstk33		&		**	10169		&			\textbf{10097}	&		*	21800			&			\textbf{21508}	&		**	34560			&			\textbf{34178}	&	*		56639			&			\textbf{54860}		&		*	80237			&			\textbf{78132}		&		+	111075			&			\textbf{108505}	\\	
	whitaker3		&		*	127			&			\textbf{126}	&		*	383				&			\textbf{380}	&		+	668				&			\textbf{656}	&	**		1150			&			\textbf{1093}		&		**	1754			&			\textbf{1717}		&		**	2676			&			\textbf{2567}		\\	
	crack			&		**	184			&			\textbf{183}	&		*	370				&			\textbf{362}	&		*	694				&			\textbf{678}	&	**		1160			&			\textbf{1092}		&		+	1815			&			\textbf{1707}		&		+	2717			&			\textbf{2566}		\\	
	wingnodal		&		*	1710		&			\textbf{1696}	&		**	3626			&			\textbf{3572}	&		**	5588			&			\textbf{5443}	&	**		8566			&			\textbf{8422}		&		*	12384			&			\textbf{11980}		&		+	16716			&			\textbf{16134}	\\	
	fe4elt2			&		**	\textbf{130}&			130				&		*	\textbf{349}	&			349				&		+	616				&			\textbf{605}	&	**		1032			&			\textbf{1014}		&		+	1694			&			\textbf{1657}		&		*	2640			&			\textbf{2537}		\\	
	vibrobox		&		*	11308		&			\textbf{10310}	&		+	19249			&			\textbf{19199}	&		+	24923			&			\textbf{24553}	&	+		34505			&			\textbf{32167}		&		**	42432			&			\textbf{41399}		&		**	51229			&			\textbf{49521}	\\	
	bcsstk29		&		**	2853		&			\textbf{2818}	&		**	\textbf{8156}	&			8379			&		*	14813			&			\textbf{13965}	&	*		23914			&			\textbf{21768}		&		*	37309			&			\textbf{34886}		&		+	58987			&			\textbf{57054}		\\	
	4elt			&		**	139			&			\textbf{138}	&		**	329				&			\textbf{321}	&		**	555				&			\textbf{534}	&	**		989				&			\textbf{939}		&		*	1639			&			\textbf{1559}		&		**	2718			&			\textbf{2596}		\\	
	fesphere		&		**	\textbf{386}&			386				&		*	794				&			\textbf{768}	&		**	1215			&			\textbf{1152}	&	*		1881			&			\textbf{1730}		&		*	2745			&			\textbf{2565}		&		+	3968			&			\textbf{3663}		\\	
	cti				&		**	334			&			\textbf{318}	&		*	973				&			\textbf{944}	&		*	1836			&			\textbf{1802}	&	**		2990			&			\textbf{2906}		&		*	4375			&			\textbf{4223}		&		*	6346			&			\textbf{5875}			\\	
	memplus			&		**	5712		&			\textbf{5489}	&		*	\textbf{9562}	&			9584			&		**	12190			&			\textbf{11785}	&	*		13908			&			\textbf{13241}		&		*	15587			&			\textbf{14489}		&		+	17381			&			\textbf{17063}		\\	
	cs4				&		+	389			&			\textbf{367}	&		*	1003			&			\textbf{940}	&		+	1568			&			\textbf{1470}	&	**		2302			&			\textbf{2206}		&		*	3228			&			\textbf{3090}		&		*	4458			&			\textbf{4169}		\\	
	bcsstk30		&		**	6391		&			\textbf{6335}	&		+	16651			&			\textbf{16622}	&		*	35037			&			\textbf{34604}	&	*		73118			&			\textbf{71234}		&		*	119316			&			\textbf{115770}		&		*	180243			&			\textbf{173945}		\\	
	bcsstk31		&		+	2769		&			\textbf{2701}	&		*	7512			&			\textbf{7444}	&		**	13608			&			\textbf{13417}	&	*		24821			&			\textbf{24277}		&		**	39455			&			\textbf{38086}		&		**	61327			&			\textbf{60528}		\\	
	fepwt			&		**	342			&			\textbf{340}	&		**	712				&			\textbf{705}	&		**	1454			&			\textbf{1442}	&	*		2844			&			\textbf{2806}		&		*	\textbf{5637}	&			5758				&		**	8648			&			\textbf{8454}	\\	
\hline
	bcsstk32		&		*	\textbf{4667}&			4667			&		*	\textbf{9440}	&			9538			&		*	21800			&			\textbf{21490}	&	**		37701			&			\textbf{37673}		&		*	63382			&			\textbf{61144}		&		*	98842			&			\textbf{95199}	\\	
	febody			&		+	266			&			\textbf{262}	&		*	\textbf{649}	&			671				&		*	\textbf{1100}	&			1156			&	*		\textbf{1910}	&			1931				&		*	\textbf{3106}	&			3202				&		*	\textbf{5212}	&			5282			\\	
	t60k			&		*	84			&			\textbf{75}		&		*	220				&			\textbf{211}	&		*	483				&			\textbf{465}	&	+		891				&			\textbf{849}		&		*	1466			&			\textbf{1391}		&		**	2297			&			\textbf{2211}	\\	
	wing			&		*	851			&			\textbf{787}	&		*	1793			&			\textbf{1666}	&		*	2720			&			\textbf{2589}	&	*		4203			&			\textbf{4131}		&		*	6217			&			\textbf{5902}		&		+	8534			&			\textbf{8132}	\\	
	brack2			&		**	731			&			\textbf{708}	&		*	3121			&			\textbf{3038}	&		*	7363			&			\textbf{7269}	&	**		12177			&			\textbf{11983}		&		**	18236			&			\textbf{17798}		&		*	27442			&			\textbf{26557}\\	
	finan512		&		**	\textbf{162}&			162				&		*	\textbf{324}	&			324				&		*	\textbf{648}	&			648				&	**		\textbf{1296}	&			1296				&		*	\textbf{2592}	&			2592				&		**	10862			&			\textbf{10560}		\\	
	fetooth			&		*	3893		&			\textbf{3823}	&		*	\textbf{7096}	&			7103			&		*	\textbf{11953}	&			12060			&	+		\textbf{18227}	&			18283				&		*	26517			&			\textbf{25977}		&		+	37079			&			\textbf{35980}		\\	
	ferotor			&		+	2103		&			\textbf{2045}	&		**	\textbf{7461}	&			7694			&		**	13283			&			\textbf{13165}	&	+		21249			&			\textbf{20773}		&		**	33266			&			\textbf{32783}		&		**	49079			&			\textbf{47461}	\\	
	598a			&		*	2426		&			\textbf{2388}	&		*	\textbf{8131}	&			8197			&		*	\textbf{16491}	&			16594			&	*		\textbf{26838}	&			27009				&		**	\textbf{40471}	&			40962				&		*	59445			&			\textbf{59098}	\\	
	feocean			&		**	468			&			\textbf{387}	&		*	1914			&			\textbf{1878}	&		+	\textbf{4270}	&			4538			&	*		\textbf{8447}	&			8507				&		+	\textbf{13673}	&			13767				&		** \textbf{21774}	&			21854	\\	
	144				&		*	6604		&			\textbf{6479}	&		*	16162			&			\textbf{15345}	&		**	26266			&			\textbf{25818}	&	*		\textbf{39195}	&			39352				&		*	58702			&			\textbf{58126}		&		*	82904			&			\textbf{81145}			\\	
	wave			&		*	8812		&			\textbf{8682}	&		**	\textbf{17616}	&			17950			&		+	\textbf{30375}	&			31697			&	+		44783			&			\textbf{44711}		&		+	\textbf{64646}	&			64860				&		*	89332			&			\textbf{88863}	\\	
	m14b			&		*	3871		&			\textbf{3826}	&		**	\textbf{13296}	&			\textbf{13403}	&		*	\textbf{26657}	&			27066			&	*		\textbf{44013}	&			44330				&		*	69072			&			\textbf{67770}		&		+	102393			&			\textbf{101551}		\\	
	auto			&		+	10329		&			\textbf{10042}	&		*	28051			&			\textbf{27790}	&		*	\textbf{47321}	&			48442			&	*		\textbf{79741}	&			81339				&		*	126146			&			\textbf{124991}		&		**	179095			&			\textbf{175975}		\\	
	\hline
\end{tabular}
\caption{Walshaw Benchmark with $\epsilon=1$ \%. * Expansion$^*$, ** Expansion$^{*2}$, + InnerOuter.}
\label{tab:walshaw1}
}\end{center}
\end{table}

\begin{table}[p]
\begin{center}
{\small
\begin{tabular}{|l|r|r|r|r|r|r|r|r|r|r|r|r|}\hline

 Graph  & \multicolumn{2}{|c|}{2} & \multicolumn{2}{|c|}{4} & \multicolumn{2}{|c|}{8} & \multicolumn{2}{|c|}{16} & \multicolumn{2}{|c|}{32} & \multicolumn{2}{|c|}{64}\\

\hline 
 3elt & 	 ** \textbf{87} & 	 87 		& 	 + 200 				& 	 \textbf{198} 		& 	 * 343 			& 	 \textbf{336} 			& 	+ 584 			& 	 \textbf{565} 	& 	 * 1010 			& 	 \textbf{958} 		& 	 + 1622 			& 	 \textbf{1542}  \\ 
 add20 & 	 + 619 & 	 \textbf{576} 		& 	 * 1179 			& 	 \textbf{1158} 		& 	 ** 1790		& 	 \textbf{1690} 			& 	+ 2161 			& 	 \textbf{2095} 	& 	 ** 2559 			& 	 \textbf{2493} 		& 	 +\textbf{3058} 	& 	 3152  \\ 
 data & 	 + 193 & 	 \textbf{185} 		& 	 ** 380 			& 	 \textbf{378} 		& 	 + 665 			& 	 \textbf{650} 			& 	** 1157			& 	 \textbf{1133} 	& 	 ** 1912 			& 	 \textbf{1802} 		& 	 * 2936 			& 	 \textbf{2809}  \\ 
 uk & 	 ** \textbf{18} & 	 18 			& 	 + 42 				& 	 \textbf{40} 		& 	 ** 82 			& 	 \textbf{81} 			& 	* 151 			& 	 \textbf{148} 	& 	 * 265 				& 	 \textbf{251} 		& 	 ** 440 			& 	 \textbf{414}  \\ 
 add32 & 	+ \textbf{10} & 	 10 		& 	 **\textbf{33}		& 	 33 				& 	 ** \textbf{66}	& 	 66			 			& 	+ 124 			& 	 \textbf{117} 	& 	 + 222 				& 	 \textbf{212} 		& 	 * \textbf{494} 	& 	 624  \\ 
 bcsstk33 & 	 + \textbf{10064}& 10064	& 	 ** 21195			& 	 \textbf{21035} 	& 	 + 34386 		& 	 \textbf{34078} 		& 	** 56262 		& 	 \textbf{54510} & 	 ** 80001			& 	 \textbf{77672} 	& 	 + 110822 		& 	 \textbf{107012}  \\ 
 whitaker3 & 	 + \textbf{126} & 126 		& 	 ** 384 			& 	 \textbf{378} 		& 	 ** 665 		& 	 \textbf{655} 			& 	+ 1138 			& 	 \textbf{1092} 	& 	 * 1753 			& 	 \textbf{1686} 		& 	 + 2655 			& 	 \textbf{2535}  \\ 
 crack & 	 + \textbf{182} & 	 182 		& 	 * \textbf{360}		& 	 360 				& 	 * 678 			& 	 \textbf{676} 			& 	+ 1126 			& 	 \textbf{1082} 	& 	 * 1782 			& 	 \textbf{1679} 		& 	 + 2670 			& 	 \textbf{2553}  \\ 
 wingnodal & 	** 1682 & 	 \textbf{1680} 	& 	 * \textbf{3565}	& 	 3566 				& 	 + 5430			& 	 \textbf{5401} 			& 	+ 8451 			& 	 \textbf{8316} 	& 	 * 12277 			& 	 \textbf{11938} 	& 	 + 16702 			& 	 \textbf{15971}  \\ 
 fe4elt2 & 	 + \textbf{130} & 	 130 		& 	 + 349 				& 	 \textbf{343} 		& 	 ** 608			& 	 \textbf{598} 			& 	** 1015			& 	 \textbf{1007} 	& 	 ** 1681 			& 	 \textbf{1633} 		& 	 ** 2617 			& 	 \textbf{2527}  \\ 
 vibrobox & 	** 11188 & 	 \textbf{10310} & 	 ** 19107			& 	 \textbf{18778} 	& 	 ** 24531 		& 	 \textbf{24171} 		& 	+ 34189 		& 	 \textbf{31516} & 	 * 42650 			& 	 \textbf{39592} 	& 	 + 50183 			& 	 \textbf{49123}  \\ 
 bcsstk29 & 	 +\textbf{2818} & 	 2818 	& 	 * 8153 			& 	 \textbf{8045} 		& 	 + 14437 		& 	 \textbf{13817} 		& 	+ 23532 		& 	 \textbf{21410} & 	 ** 37015 			& 	 \textbf{34407} 	& 	 + 58738 			& 	 \textbf{55366}  \\ 
 4elt & 	 + 138 & 	 \textbf{137} 		& 	 ** 320 			& 	 \textbf{319} 		& 	 + 536 			& 	 \textbf{523} 			& 	+ 953 			& 	 \textbf{914} 	& 	 * 1624 			& 	 \textbf{1537} 		& 	 + 2715 			& 	 \textbf{2581}  \\ 
 fesphere & 	+ \textbf{384} & 	 384 	& 	 * 796 				& 	 \textbf{764} 		& 	 + 1217 		& 	 \textbf{1152} 			& 	* 1851 			& 	 \textbf{1706} 	& 	 * 2719 			& 	 \textbf{2477} 		& 	 * 3767 			& 	 \textbf{3547}  \\ 
 cti & 	 + \textbf{318} & 	 318 			& 	 ** 927 			& 	 \textbf{917} 		& 	 * 1773 		& 	 \textbf{1716} 			& 	* 2895 			& 	 \textbf{2778} 	& 	 * 4263 			& 	 \textbf{4132} 		& 	 * 6207 			& 	 \textbf{5763}  \\ 
 memplus & 	 + 5532 & 	 \textbf{5355} 		& 	 * 9953 			& 	 \textbf{9418} 		& 	 + 12239		& 	 \textbf{11628} 		& 	+ 13755			& 	 \textbf{13237} & 	 * 15432			& 	 \textbf{14350} 	& 	 + 17870 			& 	 \textbf{17002}  \\ 
 cs4 & 	 + 383 & 	 \textbf{362} 			& 	 * 1001 			& 	 \textbf{936} 		& 	 ** 1542		& 	 \textbf{1470} 			& 	+ 2237 			& 	 \textbf{2126} 	& 	 + 3164 			& 	 \textbf{3048} 		& 	 * 4397 			& 	 \textbf{4169}  \\ 
 bcsstk30 & 	 + \textbf{6251} & 6251		& 	 *\textbf{16528}	& 	 16577 				& 	 ** \textbf{34505}& 	 34559			 	& 	* 72618 		& 	 \textbf{70278} & 	 * 118106 			& 	 \textbf{114005} 	& 	 + 179278 		& 	 \textbf{171727}  \\ 
 bcsstk31 & 	 + \textbf{2676} & 2676		& 	 **\textbf{7209}	& 	 7258 				& 	 * 13253 		& 	 \textbf{13246} 		& 	* 24365 		& 	 \textbf{23504} & 	 * 38817 			& 	 \textbf{37459} 	& 	 ** 60577 			& 	 \textbf{58667}  \\ 
 fepwt & 	 + \textbf{340} & 	 340 		& 	 + 705 				& 	 \textbf{704} 		& 	 + \textbf{1418}& 	 1421 					& 	* 2789 			& 	 \textbf{2784} 	& 	 + 5603				&
 \textbf{5606} 		& 	 ** 8630 			& 	 \textbf{8346}  \\ \hline 
 bcsstk32 & 	+ \textbf{4667} & 	 4667 	& 	 + \textbf{8805}	& 	 9533 				& 	 + \textbf{20992}& 	 21307 					& 	+ \textbf{36628}& 	 37204 			& 	 * 62639 			& 	 \textbf{59824} 	& 	 ** 97535 			& 	 \textbf{92690}  \\ 
 febody & 	 + 265 & 	 \textbf{262} 		& 	 * \textbf{613}		& 	 668 				& 	 * \textbf{1055}& 	 1094 					& 	* \textbf{1798} & 	 1903 			& 	 + \textbf{2928}	& 	 3086 				& 	 * \textbf{4997} 	& 	 5212  \\ 
 t60k & 	 + 74 & 	 \textbf{71} 		& 	 * 211 				& 	 \textbf{207} 		& 	 * 470 			& 	 \textbf{454} 			& 	* 875 			& 	 \textbf{822} 	& 	 + 1443 			& 	 \textbf{1391} 		& 	 ** 2272 			& 	 \textbf{2198}  \\ 
 wing & 	 ** 840 & 	 \textbf{774} 		& 	 * 1761 			& 	 \textbf{1636} 		& 	 * 2661 		& 	 \textbf{2551} 			& 	** 4144			& 	 \textbf{4015} 	& 	 * 6107 			& 	 \textbf{5832} 		& 	 ** 8340	 	& 	 \textbf{8043}  \\ 
 brack2 & 	 685 & 	 \textbf{684} 			& 	 * \textbf{2840} 	& 	 2864 				& 	 * 7105 		& 	 \textbf{6994} 			& 	* \textbf{11687}& 	 11741 			& 	 * 17815 			& 	 \textbf{17649} 	& 	 + 26755 			& 	 \textbf{26366}  \\ 
 finan512 &  + \textbf{162} & 	 162 		& 	 + \textbf{324}  	& 	 324				& 	 + \textbf{648}	&	 	 648 				& 	* \textbf{1296} & 	 1296 			& 	 * \textbf{2592} 	& 	 2592 				& 	 ** 10944 			& 	 \textbf{10560}  \\ 
 fetooth & 	 * 3807 & 	 \textbf{3792} 		& 	 + \textbf{6947} 	& 	 7081 				& 	 * \textbf{11562}& 	 11957 					& 	* \textbf{17678}& 	 18093 			& 	 * 25884 			& 	 \textbf{25624} 	& 	 * 36178 			& 	 \textbf{35830}  \\ 
 ferotor & 	 ** \textbf{1964} & 	 1965 	& 	 + \textbf{7263} 	& 	 7636 				& 	 ** \textbf{12798}& 	12862 				& 	+ \textbf{20404}& 	 20521 			& 	 + 32155 			& 	 \textbf{31763} 	& 	 * 47808 			& 	 \textbf{47049}  \\ 
 598a & 	 * 2373 & 	 \textbf{2367} 		& 	 * \textbf{7963}	& 	 7978 				& 	 * 16079 			& 	 \textbf{16031} 	& 	* \textbf{25960} & 	 26257 			& 	 ** \textbf{39792} 	& 	 40718 				& 	 \textbf{58430} 	& 	 \textbf{58454}  \\ 
 feocean & 	 + \textbf{311} & 	 311 		& 	 + 1706				& 	 \textbf{1704} 		& 	 * \textbf{3976} 	& 	 4019 				& 	** 8004			 & 	 \textbf{7838} 	& 	 ** 13196 			& 	 \textbf{12746} 	& 	 * \textbf{21060} & 	 21854  \\ 
 144 & 	+ 6512 & 	 \textbf{6438} 			& 	 + 15555 			& 	 \textbf{15250} 	& 	 ** \textbf{25529} & 	 25611 				& 	** 38701 		 & 	 \textbf{38478} & 	 ** 57561 			& 	 \textbf{57354} 	& 	 * 80981 			& 	 \textbf{80767}  \\ 
 wave & 	* 8699 & 	 \textbf{8616} 		& 	 * \textbf{16947} 	& 	 17407 				& 	 ** \textbf{29022} & 	 29776 				& 	* \textbf{43168} & 	 43791 			& 	 + \textbf{62766} 	& 	 63675 				& 	 + \textbf{87272} & 	 87957  \\ 
 m14b & 	* 3833 & 	 \textbf{3823} 		& 	 * \textbf{13131} 	& 	 13285 				& 	 * \textbf{26044} 	& 	 26153 				& 	* \textbf{42942} & 	 43962 			& 	 * \textbf{67272} 	& 	 67551 				& 	 + \textbf{100112}& 	 101019  \\ 
 auto & 	** 9806 & 	 \textbf{9782} 		& 	 + \textbf{26343} 	& 	 26509 				& 	 ** \textbf{45703} & 	 48263 				& 	** \textbf{77461}& 	 80495 			& 	 * \textbf{123442} 	& 	 124251 			& 	 ** 175520 		& 	 \textbf{174904}  \\ 
 \hline

\end{tabular}
}
 \end{center} 
\caption{Walshaw Benchmark with $\epsilon=3$ \%. * Expansion$^*$, ** Expansion$^{*2}$, + InnerOuter.}
\label{tab:walshaw3}
\end{table}
\end{landscape}
\begin{landscape}
\begin{table}[p]
\begin{center}{\small
\begin{tabular}{|l|r|r|r|r|r|r|r|r|r|r|r|r|}\hline

Graph  & \multicolumn{2}{|c|}{2} & \multicolumn{2}{|c|}{4} & \multicolumn{2}{|c|}{8} & \multicolumn{2}{|c|}{16} & \multicolumn{2}{|c|}{32} & \multicolumn{2}{|c|}{64}\\

\hline 
3elt 		& 	** \textbf{87} 		& 	 87 			& 	** 199 				& 	 \textbf{197} 	& 	+ 339 				& 	 \textbf{330} 		&  	+ 581 			& 	 \textbf{560} 		& 	** 1001 			& 	 \textbf{950} 		& 	 ** 1615 				& 	 \textbf{1539}  \\ 
add20 		& 	** 579 				& 	 \textbf{550} 	& 	** 1179				& 	 \textbf{1157} 	& 	+ 1744 				& 	 \textbf{1675} 		&   + 2150 			& 	 \textbf{2081} 		& 	* 2560				& 	 \textbf{2463} 		& 	 + \textbf{3054} 		& 	 3152  \\ 
data 		& 	* 188 				& 	 \textbf{181} 	& 	** 374				& 	 \textbf{368} 	& 	** 650 				& 	 \textbf{628} 		& 	+ 1147 			& 	 \textbf{1086} 		& 	* 1888	 			& 	 \textbf{1777} 		& 	 + 2910 				& 	 \textbf{2798}  \\ 
uk 			& 	** \textbf{18} 		& 	 18 			& 	+ 41 				& 	 \textbf{40} 	& 	** 81 				& 	 \textbf{78} 		& 	+ 152 			& 	 \textbf{139} 		& 	** 262 				& 	 \textbf{246} 		& 	 + 437 					& 	 \textbf{410}  \\ 
add32 		& 	** \textbf{10}	 	& 	 10			 	& 	** \textbf{33} 		& 	 33 			& 	** 66 				& 	 \textbf{65} 		& 	+ 124 			& 	 \textbf{117} 		& 	+ 222 				&	  \textbf{212} 		& 	 ** \textbf{494} 		& 	 624  \\ 
bcsstk33 	& 	** \textbf{9914} 	& 	 9914 			& 	* 20614				& 	 \textbf{20584} & 	+ 34190 			& 	 \textbf{33938} 	& 	** 55868		& 	 \textbf{54323} 	& 	+ 79530 			& 	 \textbf{77163} 	& 	 * 110822 				& 	 \textbf{106886}  \\ 
whitaker3 	& 	** \textbf{126} 	& 	 126 			& 	** 382 				& 	 \textbf{378} 	& 	** 665 				& 	 \textbf{650} 		& 	* 1130 			& 	 \textbf{1084} 		& 	* 1737	 			& 	 \textbf{1686} 		& 	 + 2655 				& 	 \textbf{2535}  \\ 
crack 		& 	** \textbf{182} 	& 	 182 			& 	** \textbf{360} 	& 	 360 			& 	** 679 				& 	 \textbf{667} 		& 	+ 1122 			& 	 \textbf{1080} 		& 	* 1755  			& 	 \textbf{1679} 		& 	 + 2651 				& 	 \textbf{2548}  \\ 
wingnodal 	& 	* 1676 				& 	 \textbf{1668}  & 	* 3545				& 	 \textbf{3536}  & 	+ 5376 				& 	 \textbf{5350} 		& 	+ 8388 			& 	 \textbf{8316} 		& 	** 12252 			& 	 \textbf{11879} 	& 	 ** 16595 				& 	 \textbf{15873}  \\ 
fe4elt2 	& 	** \textbf{130} 	& 	 130 			& 	** 349 				& 	 \textbf{335}   & 	* 599 				& 	 \textbf{583} 		& 	** 1015 		& 	 \textbf{991} 		& 	** 1660 			& 	 \textbf{1633} 		& 	 + 2609 				& 	 \textbf{2516}  \\ 
vibrobox 	& 	** 11188 			& 	 \textbf{10310} & 	** 18958 			& 	 \textbf{18778} & 	* 24121 			& 	 \textbf{23930} 	& 	* 33760 		& 	 \textbf{31235} 	& 	** 42269 			& 	 \textbf{39592} 	& 	 + 49552 				& 	 \textbf{48200}  \\ 
bcsstk29 	& 	** \textbf{2818} 	& 	 2818 			& 	+ 8055 				& 	 \textbf{7942}  & 	* 14009 			& 	 \textbf{13614} 	& 	+ 23131 		& 	 \textbf{20924} 	& 	** 36633 			& 	 \textbf{33818} 	& 	 * 58183 				& 	 \textbf{54935}  \\ 
4elt 		& 	** \textbf{137} 	& 	 137 			& 	* 319 				& 	 \textbf{315} 	& 	* 526 				&	 	 \textbf{516} 	& 	** 946 			& 	 \textbf{902} 		& 	* 1590 				& 	 \textbf{1532} 		& 	 * 2675 				& 	 \textbf{2565}  \\ 
fesphere 	& 	** \textbf{384} 	& 	 384 			& 	** 784 				& 	 \textbf{764} 	& 	+ 1217 				& 	 \textbf{1152} 		& 	* 1840 			& 	 \textbf{1692} 		& 	** 2709 			& 	 \textbf{2477} 		& 	 + 3945 				& 	 \textbf{3547}  \\ 
cti 		& 	** \textbf{318} 	& 	 318 			& 	+ \textbf{891} 		& 	 897 			& 	* 1737				& 	 \textbf{1716} 		& 	** 2885 		& 	 \textbf{2725} 		& 	* 4242 				& 	 \textbf{4037} 		& 	 + 6010 				& 	 \textbf{5684}  \\ 
memplus 	& 	** 5528 			& 	 \textbf{5267} 	& 	+ 9489 				& 	 \textbf{9299} 	& 	** 12091 			& 	 \textbf{11555} 	& 	+ 13701 		& 	 \textbf{13078} 	& 	** 15362 			& 	 \textbf{14249} 	& 	 + 17632 				& 	 \textbf{16662}  \\ 
cs4 		& 	* 373 				& 	 \textbf{356} 	& 	* 990 				& 	 \textbf{936} 	& 	** 1542 			& 	 \textbf{1470} 		& 	+ 2237 			& 	 \textbf{2126} 		& 	* 3141	 			& 	 \textbf{2995} 		& 	 ** 4364 				& 	 \textbf{4116}  \\ 
bcsstk30 	& 	** \textbf{6251} 	& 	 6251 			& 	** \textbf{16316} 	& 	 16417 			& 	** \textbf{34391} 	& 	 34559 				& 	+ 72087 		& 	 \textbf{70043} 	& 	** 117512 			& 	 \textbf{113321} 	& 	 ** 177303 				& 	 \textbf{170591}  \\ 
bcsstk31 	& 	** \textbf{2676} 	& 	 2676 			& 	** \textbf{7118} 	& 	 7223 			& 	* 13104				& 	 \textbf{13058} 	& 	+ 24062 		& 	 \textbf{23254} 	& 	** 38279 		 	& 	 \textbf{37459} 	& 	 ** 60257 				& 	 \textbf{57534}  \\ 
fepwt 		& 	** \textbf{340} 	& 	 340 			& 	** \textbf{700} 	& 	 704 			& 	** \textbf{1406} 	& 	 1411 				& 	+ \textbf{2773} & 	 2778 				& 	+ \textbf{5525} 	& 	 5606 				& 	 + 8582 				& 	 \textbf{8310}  \\
		       \hline	
bcsstk32 	& 	** \textbf{4667} 	& 	 4667 			& 	+ \textbf{8539} 	& 	 9052			& 	** 20568 			& 	 \textbf{20099} 	& 	+ \textbf{35962}& 	35990 				& 	** 61021 			& 	 \textbf{59824} 	& 	 ** 96032 				& 	 \textbf{91006}  \\ 
febody 		& 	** 263 				& 	 \textbf{262} 	& 	* \textbf{599} 		& 	 629 			& 	* \textbf{1055}		& 	 1072 				& 	* \textbf{1786} &  	 1815 				& 	+ \textbf{2863} 	& 	 3086 				& 	 + \textbf{4897} 		& 	 5043  \\ 
t60k 		& 	** 69 				& 	 \textbf{65} 	& 	+ 206 				& 	 \textbf{196} 	& 	* 469 				& 	 \textbf{454} 		& 	* 865 			& 	 \textbf{818} 		& 	** 1436 			& 	 \textbf{1376} 		& 	 ** 2263 				& 	 \textbf{2168}  \\ 
wing 		& 	+ 826 				& 	 \textbf{770} 	& 	** 1734 			& 	 \textbf{1636} 	& 	* 2632 				& 	 \textbf{2551} 		& 	* 4106 			& 	 \textbf{4015} 		& 	* 6063  			& 	 \textbf{5806} 		& 	 + 8300 				& 	 \textbf{7991}  \\ 
brack2 		& 	** \textbf{660} 	& 	 660 			& 	** \textbf{2739} 	& 	 2755 			& 	* \textbf{6776} 	& 	 6883 				& 	+ \textbf{11557}& 	 11558 				& 	* 17617 			& 	 \textbf{17529} 	& 	 + 26555 				& 	 \textbf{26281}  \\ 
finan512 	& 	** \textbf{162} 	&	 162 			& 	** \textbf{324} 	& 	 324 			& 	** \textbf{648}  	& 	 648 				& 	* \textbf{1296} & 	 1296 				& 	** \textbf{2592} 	& 	 2592 				& 	 ** 10909 				& 	 \textbf{10560}  \\ 
fetooth 	& 	** 3785 			& 	 \textbf{3773} 	& 	* \textbf{6863} 	& 	 7027 			& 	+ \textbf{11498} 	& 	 11957  			& 	* \textbf{17509}& 	 18090 				& 	* 25641 			& 	 \textbf{25624} 	& 	 + 35795 				& 	 \textbf{35476}  \\ 
ferotor 	& 	** \textbf{1955} 	& 	 1957 			& 	+ \textbf{7031} 	& 	 7520 			& 	* \textbf{12643} 	& 	 12678 				& 	** \textbf{20098}& 	 20263 				& 	+ 31611 			& 	 \textbf{31576} 	& 	 + 47186 				& 	 \textbf{46608}  \\ 
598a 		& 	** 2344 			& 	 \textbf{2336} 	& 	+ \textbf{7837} 	& 	 7978 			& 	** \textbf{15794} 	& 	 16031 				& 	* \textbf{25782} & 	 26257 				& 	** \textbf{39478} 	& 	 40179 				& 	 ** \textbf{58180} 		& 	 58307  \\ 
feocean 	& 	** \textbf{311} 	& 	 311 			& 	** \textbf{1688} 	& 	 1704 			& 	* \textbf{3952} 	& 	 4019 				& 	+ \textbf{7671}  & 	 7838 				& 	+ 12953 			& 	 \textbf{12746} 	& 	 + \textbf{20660} 		& 	 21784  \\ 
144 		& 	* 6502 				& 	 \textbf{6362}	& 	+ 15313 			& 	 \textbf{15250} & 	** \textbf{25529} 	& 	 25611 				& 	+ \textbf{38182} & 	 38478 				& 	+ \textbf{57202} 	&	 	 57354 			& 	 + 80653 				& 	 \textbf{80257}  \\ 
wave 		& 	** 8613				& 	 \textbf{8563} 	& 	+ \textbf{16780} 	& 	 17306 			& 	+ \textbf{28753} 	& 	 29776 				& 	+ \textbf{42810} & 	 43791 				& 	** \textbf{62382} 	& 	 63675 				& 	 ** \textbf{86867} 		& 	 87654  \\ 
m14b 		& 	** 3844				& 	 \textbf{3802} 	& 	* \textbf{13124} 	& 	 13285 			& 	** \textbf{25701} 	& 	 26153 				& 	+ \textbf{42644} & 	 43747 				& 	* \textbf{66845} 	& 	 67551 				& 	 + \textbf{99460} 		& 	 100183  \\ 
auto 		& 	* 9587 				& 	 \textbf{9450} 	& 	+ \textbf{25805} 	& 	 26097			& 	** \textbf{44915} 	& 	 48174 				& 	+ \textbf{76500} & 	 80495 				& 	** \textbf{121988}	& 	 124251 			& 	 ** \textbf{174173}		& 	 174904  \\ 
\hline

\end{tabular}}
\end{center} 
\caption{Walshaw Benchmark with $\epsilon=5$ \%. * Expansion$^*$, ** Expansion$^{*2}$, + InnerOuter.}
\label{tab:walshaw5}
\end{table}

\end{landscape}

\end{appendix}
\end{document}